\documentclass[a4paper]{article}

\usepackage[mathletters]{ucs}
\usepackage[utf8x]{inputenc}
\usepackage{amsmath}
\usepackage{amssymb}
\usepackage{stmaryrd}
\usepackage{zed-csp}
\usepackage{array}

\usepackage{wasysym}

\newdimen\eventwidth
\def\event#1{\def\parm{#1} \setbox0\hbox{$#1$}\eventwidth=\wd0
\advance\eventwidth by 1.2em \mathop{\hbox to \eventwidth
{\rightarrowfill}}\limits \ifx\parm\empty\else ^{\box0}\fi}

\newcommand{\eq}{\; \textrm{\Large  =}\;}

\DeclareMathOperator{\pow}{\ensuremath{\mathbb{P}}}
\newcommand{\bpre}{\mathrel{\,\rule[-0.15em]{0.15em}{1em}\,}}

\newcommand{\asemi}{  {\mathrel {\rm \bf \, ; \!}  }  }
\newcommand{\Guard}{\Longrightarrow}
\newcommand{\unpack}{\mathop{\sim}}

\newcommand{\lbb}{\llparenthesis}
\newcommand{\rbb}{\rrparenthesis}

\newcommand{\drangle}{\rangle \! \rangle}
\newcommand{\lboldb}{\pmb{\boldsymbol{(}}}
\newcommand{\rboldb}{\pmb{\boldsymbol{)}}}

\newcommand{\pref}{\rightslice}

\renewcommand{\[}{\begin{argue}}
\renewcommand{\]}{\end{argue}}

\newcommand{\bguard}{\mathrel{\rightarrowtriangle}}

\newcommand{\q}[1]{\text{``#1''}}

\def\IF{{ \; {\rm \bf if} \;}}
\def\THEN{ \; {\rm \bf then} \; }
\def\ELSE{ \; {\rm \bf else} \; }
\def\END{ \; {\rm \bf end} \; }
\def\WHILE{\; {\rm \bf while} \; }
\def\DO{ \; {\rm \bf do} \; }

\newcommand{\maths}{$\begin{array}{l}}
\newcommand{\xmaths}{\end{array}$}

\newcommand{\Maths}{$$\begin{array}{l}}
\newcommand{\Xmaths}{\end{array}$$}

\newcommand*{\rules}{$\begin{array}{lll}}
\newcommand*{\xrules}{\end{array}$}

\newcommand{\Rules}{$$\begin{array}{lcc}}
\newcommand{\Xrules}{\end{array}$$}

\numberwithin{equation}{section}

\newcommand{\eqns}{\begin{align*}}
\newcommand{\eeqns}{\end{align*}}

\usepackage{appendix}

\newcommand{\rightq}{$'' $}

\newcommand{\Exp}{{\bf \rm E}}

\newcommand{\bs}{\backslash}
\newcommand{\card}{{\rm card}}
\renewcommand{\t}{\hspace{10pt}}  

\title{Bunch theory: applications, axioms and models}

\author{Bill Stoddart, Frank Zeyda, Steve Dunne}

\begin{document}

\maketitle

\begin{abstract} 
In his book {\em A practical theory of programming} \cite{ECRH93}, Eric Hehner proposes and applies a remarkably radical reformulation of set theory, in which the collection and packaging of elements are seen as separate activities. This provides for unpackaged collections, referred to as “bunches”.  Bunches allow us to reason about non-determinism at the level of terms, and, very remarkably, allow us to reason about the conceptual entity “nothing”, which is just an empty bunch (and very different from an empty set). This eliminates mathematical “gaps” caused by undefined terms. We compare the use of bunches with other approaches to this problem, and we illustrate the use of Bunch Theory in formulating program semantics combining non-deterministic, preferential, and probabilistic choice to provide a guarded command language whose exceptional expressivity we illustrate with a  short case study.  We show how an existing axiomatisation of set theory can be extended to incorporate bunches, and we provide and validate a model. 
\end{abstract}

\tableofcontents
\section{Introduction}

Set theory deals with elements as singular entities. A set may contain elements, and is itself an element that can belong to another, more deeply structured, set. When forming a set from its elements we simultaneously perform two actions: collection and packaging. However, it is beneficial to think of these as two orthogonal activities.  The theory that results from this approach we will refer to as bunch theory, following Hehner, who first suggested it. 

Bunch theory provides benefits in terms of preserving homogeneity, accommodating what would otherwise be undefined terms, and describing non-determinism - a key concept in software development.

Hehner provides an axiomatisation for Bunch Theory and builds set theory on top of this.  Our approach will be to take an existing axiomatisation of set theory and add additional axioms to incorporate bunches.

The base formalism we build on is that used in the B-Method of J-R Abrial \cite{JRA96}. This gives us first order predicate logic with equality (FOPL) and a typed set theory. The B-Method provides us with the ability to construct basic sets, 
including the integers. Such sets are termed “maximal” since they are not subsets of any larger set. From any existing maximal sets we are able to produce new maximal sets by the use of set product and power-set operations. Each maximal set corresponds to a type. 

Following Hehner \cite{ECRH93}, we give a mathematical meaning to the contents of a set, which we call a bunch.  We write $\unpack A$ (“unpackage $A$”) for the contents of set $A$, thus $\unpack \{1,2\} \, = \, 1,2$. The comma used in a bunch extension is now a mathematical operator, called bunch union. It is associative and commutative, and its precedence is just below that of the expression connectives.\footnote{A full list of precedences is given in Appendix A.} It is associated with set union via the rule

$\{A\} \, ∪ \, \{B\} \, = \, \{A,B\}$ 

A bunch $A$ is a sub-bunch of $B$ if each element of $A$ is an element of $B$. We write this as $A \, : \, B$ (“$A$ is part of $B$”). Sub-bunches are related to subsets by the rule:

$A \, : \, B \,  \, ⇔ \,  \, \{A\} \, ⊆ \, \{B\}$

We define the bunch $null$ to  be the contents of the empty set. $null$ literally represents nothing, a concept completely missing from other theories but which for us plays an essential rôle.

 The introduction of bunches affects our logic. In FOPL all terms represent values, and it is possible to prove $∃x\bullet x=1/0$, whereas in bunch theory $1/0=null$. In \cite{arXivLogic} we propose a weakening of FOPL which we call  FOPLN.  FOPLN introduces a definedness predicate $δ(E)$ to say that $E$ is an element. We retain propositional logic  and can still prove almost all predicate logic rules from our logic toolbox, the only exceptions being that one point rules require a definedness premise.

Standard function application is lifted when applied to a bunch, thus if $A=x,y$, then $f(A)=f(x),f(y)$.  We take the infix syntax of binary operations as a sugared form of standard function application, so these are similarly lifted to apply pairwise, e.g. adding the bunches $0,1$ and $2,4$ yields $0+2, \, 0+4, \, 1+2, \, 1+4$. Using round brackets exclusively to express precedence (rather than sometimes using them to describe ordered pairs) and noting that bunch union (comma) has a lower precedence than $+$, we can write this as: \((0,1) \, + \, (2,4) \, = \, 2,3,4,5\). Thus we sacrifice the use of standard brackets to represent tuples: $(x,y)$ is no  longer an ordered pair, for which we use maplet notation, i.e. $x \, \mapsto  \, y$  or bold brackets: $\lboldb x,y\rboldb $. However we still use $f(x,y)$ for application  of function $f$ to a pair of arguments, and when we apply a function $f$ to a bunch of arguments $x,y$, we write this as $f((x,y))$.

 Lifting similarly applies to tuple construction, so that, e.g. $(1,2)\mapsto (3,4) \, = \, 1\mapsto 3,1\mapsto 4,2\mapsto 3,2\mapsto 4$ 

Lifting does not, however, apply to “packaging", which is the inverse of  the unpackaging operator described above. We package the bunch $1,2$ by  enclosing it in set brackets, obtaining $\{1,2\}$.

For us, as in B, a predicate is totally  different from an expression. A predicate is a formula that may be subject to proof, whereas expressions denote values. In particular our predicates {\em do not denote boolean truth values}. Accordingly, predicates are not lifted in the way  described above. 

For use with bunches the two fundamental predicates we inherit from B, set membership, and equality, must be treated in different ways.

$E \, ∈ \, S$ will be true when each element which is part of $E$ is a member of each element which is part of $S$. Thus $1,2 \,  \, ∈ \,  \, \{1,2,3\},\{2,3,4\}$ For elements $x$ and $y$ a predicate such as $x<y$ is equivalent to $x\mapsto y \, ∈ \, \_<\_$ where $\_<\_$ is the relation containing  all ordered pairs $x\mapsto y$ for which $x<y$. Thus we have $1,2 \, < \, 3,4$, but $¬(1,2 \, < \, 2,3)$.

We cannot apply this treatment to equality  since we inherit from first order predicate logic with equality the axiom $E=E$. Therefore, for example, $1,2 \, = \, 1,2$.  

The guarded bunch \cite{NH92}, $p \, \bguard  \, E$ is equal $null$, if predicate $p$ is false, and is otherwise equal to the expression $E$. 

We will see many uses of guarded bunches. One we can give immediately is to define the conditional expression, for predicate $p$ and expressions $E, \,  \, F$.

$\IF \, p \, \THEN \, E \, \ELSE \, F \, \END \,  \, \defs  \,  \, ( \, p \, \bguard  \, E \, ) \, , \, ( \, ¬p \, \bguard  \, F \, )$

 The improper bunch $⊥ \, $ is used to describe “non-termination”.  It represents the complete ignorance associated with a value provided by a computation that may have crashed. Such a failed computation might produce any value, or no value at all, and this is reflected in the improper bunch being strictly larger than any other bunch of its type, i.e. for any  proper bunch $A$, we have $A:⊥$ and $¬(A=⊥)$.  \footnote{There is a different null bunch and a different improper bunch for each type. Where we need to specify the type we do so by means of a suffix, e.g. $⊥_\num $ for the bottom bunch of type integer, but usually the type of any $null$ or improper bunch we are considering will be obvious from context.}

The preconditioned bunch $p \, \bpre  \, E$ is equal to the improper bunch $⊥$ if $p$ is false, and otherwise equal to $E$.

From any set $X$ we can construct a smaller set by set comprehensions of the form

$\{x \, | \, x \, ∈ \, X \, ∧ \, P \, \}$

All sets we consider are necessarily homogeneous.

We similarly restrict our attention to bunches which are homogeneous; these have the same type as their elements, thus $1,2$ is of type integer.

Analogously to set comprehension, we can write the bunch comprehension:

$∮ \, x \, \bullet  \, E$

Here, although the type of $x$ is not given explicitly it has to be deducible from an examination of the expression $E$. The result consists of all values that $E$ can take as $x$ ranges over the elements of its type. For example the bunch $10,20$ can be written as the bunch comprehension $∮ \, x \, \bullet  \, x:1,2 \, \bguard  \,  \, 10*x$.

When we extend set theory to include bunches, a reasonable ambition is for every  closed formula of the original theory to have the same meaning in the  extended theory. This requires bound variables  to denote elements, a property of FOPLN which we prove in \cite{arXivLogic}. For example in $∀ \, n \, \bullet  \, n \, ∈ \, \num  \, ⇒ \, \card \, \{ \, n \, \} \, = \, 1$ is true in our original set theory, and remains true in our extended theory since the bound variable $n$ denotes an element. 

\section{The structure of the rest of  the paper}

In section \ref{uses} we look at two application areas of bunch theory. Firstly we analyse the contribution it can make to theories of description due to its ability to eliminate undefined terms by representing them as $null$. Secondly, we show how use of bunches can facilitate the formulation of programming semantics, including combinations of non-deterministic,  provisional and preferential and probabilistic choice within a backtracking context.  These are illustrated with a case study  that  implements a search heuristic. 

Since the introduction of bunches into set theory is a radical departure from mathematical tradition, we justify its soundness, along with the soundness of its logic FOPLN,  by  providing a model. In section 4 we extend the axiomatisation of set theory given in \cite{JRA96} to incorporate bunch theory. In section 5 we define a model, which we validate in section 6. To simplify our validation, we keep our notation as sparse as possible, and in section 7 we formally extend our theory to include notations previously described in the informal introduction to bunch theory. In section 8 we report some results from bunch theory. In section 9 we extend our model to incorporate “improper bunches”, which are used to capture the occurrence and spread of error conditions, such as non-termination, which must be presumed when operations are executed outside their pre-conditions. In section 10 we report related work and in section 11 we give our conclusions. 

Appendix \ref{Fixpoint} provides a fixed point treatment of loops and appendix \ref{precedence} gives details of operator precedence and associativity.

The original contributions of this paper include an analysis of the contribution of bunch theory to the problem of undefined terms, a case study which illustrates the most novel aspects of program semantics, an axiomatisation based on extending set theory by giving a formal meaning to the contents of a set, the provision and validation of a model for this approach and for our logic FOPLN which demonstrates the soundness of our theories, and a semantics for the “improper bunch” which our program semantics uses to model error conditions using the total correctness approach of R Back\cite{RJRB81}, and used in the B-Method \cite{JRA96}, and in the Integrated Theories approach of  Hoare and He \cite{HH98} .

\section{Bunch theory in use} 

\label{uses}

We examine some contributions made by Bunch Theory, concentrating on three main areas: the resolution of problems associated with undefined terms, the expression of choice and randomness in programs, and the preservation of homogeneity to simplify formal descriptions.

\subsection{Bunch theory and descriptions}\label{descriptions}

Bunch theory solves many problems relating to “descriptions” that include undefined values and non-denoting terms and also values described in terms of a computation which does not terminate. For example the term $1/0$ will represent $null$, and this has well defined mathematical behaviour. We can prove $1/0 \, = \, 1/0$, which conforms with the rule for equality $E \, = \, E$ in First Order Predicate Logic with Equality. However, before we discuss the full impact of bunch theory on descriptions, let us briefly survey how troublesome the area of undefined terms has been.

In our formal notation, as in natural language, many objects are referred to
by  means of a description rather than being named directly. The problems which arise when such descriptions do not refer to anything (for example $1/0$ or "the present king of France"), or are ambiguous, have  been the source of extensive debate in mathematics and philosophy at least since the emergence of Russell's Theory of Descriptions (RTD)  \cite{BR05,WR10}. In response to the proposals of Meinong and Frege, Russell rejected approaches that attempted to directly assign some form of meaning or denotation to an invalid description (also referred to an as “undefined term”), but showed how statements that involved such terms could nevertheless be handled in a consistent way. In RTD, undefined terms are non-denoting.  They have no significance of their own account, but any assertion in which they have a primary occurrence entails an existential claim which results in the assertion being false. For example “the present king of France is bald” is false, since it entails a claim that there is a king of France;  “the King of France is non-bald” is also false, but, by negation of our first example, “it is false that the King of France is bald” is true.

In formal notations, the definite description $ι \, x \, \bullet  \, P$ commonly represents the supposedly unique value of $x$ described by predicate $P$. A common form of definite description is function application, noting that the term $f(x)$ can be expressed as  $ι \, y \, \bullet  \, x \, \mapsto  \, y \, ∈ \, f$. In a formalism in which partial and total functions are a particular form of binary relation (an approach which RTD shares with a number of current formalisms including Z, B, and VDM) the term $f(x)$ is taken to be an invalid description when there is no unique $y$ such that $x \, \mapsto  \, y \, ∈ \, f$.

In Abrial's approach, instead of undefined terms being non-denoting (as in RTD), we hold that they do denote an object of the correct type, but we do not know what it is. This provides the advantage of a formal system compatible with predicate logic with equality. In particular we have the simple equality law $E=E$. For example we can prove $1/0 \, = \, 1/0$. This is not possible in RTD, because it entails the existential claim that there is a number that is $1/0$.  However, Abrial's approach means that stating $f(x)=y$ is not enough for us to conclude $x \, ∈ \, dom(f)$. To see why suppose we have the following rule for function application:

\[f(a) \, = \, b \,  \, ⇒ \,  \, a\mapsto b \, ∈ \, f \, \]

{\em This is the function application rule in our formalism,} which includes bunches, but does not work in Z and B. Using the empty relation $\{ \, \}$ as a function to deliberately provoke an undefined term, we can show, using this rule, that something belongs to the empty set. From our reflexive equality rule we have $\{\}(a) \, = \, \{\}(a)$. Hence, from the function application rule, we immediately obtain $a \, \mapsto  \, \{\}(a) \, ∈ \, \{\}$.  Now since $\{\}a$ denotes some value (though we do not know what this is) so also must $a \, \mapsto  \, \{\}(a)$. But we have seen  $a \, \mapsto  \, \{\}(a) \, ∈ \, \{\}$: we have constructed an element that belongs to the empty set!  To avoid this problem we must formulate a rule for function application that includes information about the domain and functionality of $f$ e.g.

\[f \, ∈ \, A \, \pfun  \, B \, ∧ \, a \, ∈ \, dom(f) \, ∧ \, f(a) \, = \, b \,  \, ⇒ \,  \, a\mapsto b \, ∈ \, f \, \]

and when writing a formal text the user has to be careful to give all the information required, which is not always easy, as in a recursive definition the domain of the function might not be explicitly given \cite{SDG99}. 

Such considerations have led C B Jones and his associates to construct a three valued logic \cite{HB84}, in which predicates can be $true$, $false$, or $undefined$. In this approach we cannot fall back on classical logic, and we lose the law of the excluded middle as well as the simple law for equality. However, $f(x)=y$ does tell us that $y$ is the unique value associated with $x$ in $f$.

In the specification language Larch, and in the Isabelle/HOL proof assistant, partial functions are not available, and applications  requiring functions that change according to events are modelled using total functions but giving only limited information about them. Asserting $f(a)=b$ then provides information about $f$ without any need for other conditions to hold. To model a birthday book we might have a set that associates people with dates, and rather than the set representing a partial function, it would contain information about a total function $birthday$ (say) over part of its domain. Because $birthday$ is a total function, mapping, say from $PERSON$ to $DATE$ we can say for any person $n$ that $∃ \, d \, \bullet  \, d \, ∈ \, DATA \, ∧ \, birthday(n)=d$. We might not know the date of a person's birthday, but we have no objection to the fact that such a date exists. However, this does not fit so well with a function  $king\_of$ mapping from $COUNTRY$ to $PERSON$, as it allows $∃ \, p \, \bullet  \, p \, ∈ \, PERSON \, ∧ \, p \, = \, king\_of(france)$. I.e. there is an entailment that France has a king, though we don't know who it is.

There are also approaches which reject predicates with undefined terms as possible candidates for proof. This may be done by requiring the discharge of some preliminary well definedness proof obligations. Alternatively, it may be achieved by syntactic restrictions that require predicates to be written in a way  that ensures well definedness.\cite{AM02} 

Some of the many approaches, are described in \cite{DG96,SDG99}, and in \cite{AM02} where the authors consider the treatment of the propositions $2+3/0 \, = \, 2$ and $2+3/0 \, ≠ \, 2$ under 6 different regimes of undefinedness, none of which agree over the two results.

Let us now examine them in bunch theory. For the first we have:

\[2+3/0 \, = \, 2 \,  \,  \, ≡ \,  \,  \, 2+null \, = \, 3 \,  \, ≡ \,  \,  \, null \, = \, 3 \,  \, ≡ \,  \, false \, \]

In our formalism, the second is syntactic sugar for a set membership predicate, where $\_≠\_$ is the relation consisting of pairs of unequal integer values, and where the type $\num  \, ↔ \, \num $ of the relation is given by context.  We have:

\[2+3/0 \,  \, ≠ \,  \, 2 \,  \,  \, ≡ \,  \,  \, 2+3/0 \, \mapsto  \, 2 \,  \, ∈ \,  \, \_≠\_ \,  \,  \, ≡ \,  \,  \, null \, \mapsto  \, 2 \,  \, ∈ \,  \, \_≠\_ \,  \,  \\
≡ \,  \,  \, null \, ∈ \, \_≠\_ \,  \,  \, ≡ \,  \, true \, \]

Now consider the proposition:

\[lives\_in(son\_of(Charles)) \, = \,  \, uk \, \] 

where Charles has two sons. We think of the relation $son\_of$ as a non-deterministic function, which in this case returns a bunch of two values. The proposition will be true iff both sons live in the UK. The argument for this is that the bunch of two elements represent a non-deterministic choice, and as with non-deterministic choice in abstract command languages, we require that either choice must satisfy the given condition. 

Returning to Russell's example, we represent this as

\[king\_of(france) \,  \, ∈ \,  \, bald \,  \,  \, ≡ \,  \,  \, null \, ∈ \, bald \,  \,  \, ≡ \,  \, true \,  \, \]

thus in this case we disagree with RTD, but the truth we obtain is vacuous, like the predicate established by an infeasible operation \cite{CM90}, or a universal quantification over an empty set.

It is possible to avoid undefined terms by using relational image rather than function application. The relational image of set $S$ under relation $R$ is defined as the set of these values $y$ taken from pairs $x \, \mapsto  \, y \, ∈ \, R$ for which $x∈S$, which we write as:

\( \, R[S] \, \defs  \, \{ \, x \, , \, y \, | \, x\mapsto y \, ∈ \, R \, ∧ \, x∈S \, \bullet  \, y \, \} \, \)

and there is a close relationship with function application, since if $f(a)=b$ then $f[\{a\}] \, = \, \{b\}$. Relational image is always well defined, and allows us to capture non-existent objects via use of the empty set, e.g. $king\_of \, [ \, \{ \, france \, \} \, ] \, = \, \{ \, \}$. An interesting example of its use can be found in Version 1 of the Z Notation Standard \cite{ZBS}, a document which throughout uses relational image rather than function application. However, this technique is inevitably clumsy and was abandoned in later versions of the Standard.\footnote{In our bunch theory relational image is not defined since the value of $R[S]$ can be written as $\{ \, R(\unpack S) \, \}$.} 

Finally we note that the underlying logic for our theory is Predicate Logic with Equality. Thus we have a classical two valued logic and the simple equality rule holds, in which every expression is equal to itself. We do not require special requirements on the domain of a function to draw conclusions from a function application, as for elements $f \, , \, x \, , \, y$ we have the rules:

\[f(x)=y \, ⇒ \,  \, x \, \mapsto  \, y \, ∈ \, f \,  \\
 \,  \,  \, f(x)=y \, ⇒ \, ¬(∃ \, z \, \bullet  \, z≠y \, ∧ \, a\mapsto z \, ∈ \, f)\]
$\cent $
These allow us to define a function by its applicative effect and without additional reference to its domain. Let's see how this works in the case of the factorial function, given the following properties which characterise it as an element and describe its application. Note that we use Hehner's large equals as a low precedence equality.

\[  \,  \,  \, \cent (fact)=1 \,  \\
 \,  \,  \, fact(n) \, \eq \, \IF \, n=0 \, \THEN \, 1 \, \ELSE \, n*fact(n-1) \, \END \\
\] 

These properties suggest an operational interpretation for executing the function $fact$, but in our approach we don't have direct access to such an interpretation. We follow the B-Method in taking functions to  be mathematical rather than computational objects, and we handle non-termination at the level of operations by  means of operation pre-conditions. Here, $fact$ is particular type of set, a function from $\nat  \, → \, \nat $, and we need a formal way to relate the above description of $fact$ to this set. Furthermore, we wonder what would be a reasonable interpretation of the factorial of a negative number, for example $fact(-3)$. The interpretation our theory  provides is that $fact(-3)=null$ on the grounds that no value is described by $fact(-3)$. 

We take the usual approach of using a higher order function $F$ to formulate a fixed point equation $F(fact)=fact$ and finding $fact$ as the solution of this equation. We will be able to obtain the descriptive interpretation of $fact$ as the greatest fixed point solution, $fact=\nu F$.

\[F \, \defs  \, λ \, f \, \bullet  \, λ \, n \, \bullet  \, \IF \, n=0 \, \THEN \, 1 \, \ELSE \, n*f(n-1) \, \END \, \]

The existence of greatest and least fixed point solutions to $F(fact)=fact$ is guaranteed by the Knaster-Tarski theorem. The associated partial order, which reflects how deterministic a function is, is the inverse of set inclusion ordering, and is thus a complete lattice. Since $F$ is continuous, as well as monotonic, we will be able to derive our solutions as the limits of monotonic sequences of approximations to $fact$. To obtain the greatest fixed point $\nu F$ we use as first approximation the everywhere over-deterministic function $\{ \, \}$. Each successive approximation is obtained by applying $F$ to the previous one. Thus:

$\begin{array}{l}  \,  \, fact_0{} \, = \, \{ \, \} \,  \,  \,  \end{array}$

$\begin{array}{l}
 \,  \, fact_1{} \, = \, F(fact_0{}) \, = \,  \\
 \,  \, λ \, n \, \bullet  \, \IF \, n=0 \, \THEN \, 1 \, \ELSE \, n*\{\}(n-1) \, \END \, = \\
 \,  \, λ \, n \, \bullet  \, \IF \, n=0 \, \THEN \, 1 \, \ELSE \, null \, \END \,  \, = \,  \\
 \,  \, \{ \, n \, | \, n=0 \, \bguard  \, 1, \, ¬(n=0) \, \bguard  \, null \, \} \, = \\
 \,  \, \{ \, 0 \, \mapsto  \, 1 \, \} \end{array}$

$\begin{array}{l}
 \,  \, fact_2{} \, = \, F(fact_1{}) \, = \\
 \,  \, λ \, n \, \bullet  \, \IF \, n=0 \, \THEN \, 1 \, \ELSE \, n*\{0\mapsto 1\}(n-1) \, \END \, = \,  \,  \\
 \,  \, \{ \, n \, | \, n=0 \, \bguard  \, 1, \, ¬(n=0) \, \bguard  \, n=1 \, \bguard  \, 1 \, \} \, = \\
 \,  \, \{ \, 0 \, \mapsto  \, 1 \, , \, 1 \, \mapsto  \, 1 \, \} \,  \\
\end{array}$

Each successive approximation nibbles away at the over-deterministic part of the previous function, providing one more point where a result is known. In the limit, the result of applying $\nu F$ to any non-negative argument is known. However, application to a negative value remains overdeterministic, yielding $null$,  in accordance with our descriptive requirements.

To obtain $μF$, the weakest fixed point solution of $F(fact)=fact$, we make our first approximation $All$, defined to be the function that returns $\unpack \num $ at every  point where it is applied. In this case each successive approximation replaces one point at which the previous approximation yielded $\unpack \num $ with a point that yields a deterministic value. In the limit, the result of applying $μF$ to any non-negative argument is known. However, application to a negative value remains maximally non-deterministic, yielding $\unpack \num $. 

The reader may wonder whether we could use, as an initial approximation, a function which is everywhere non-terminating, and thus provide a computational interpretation of a function which shows for which arguments it would fail to terminate.  That would require a more complicated theory of bunches than we provide here, but see Appendix \ref{Fixpoint} for a way in which we can overcome this apparent limitation when considering the fixed point treatment of loops.\\

In \cite{JLS12} Jones et al classify attempts to deal with undefined terms into three main categories: those that use a logic adapted to undefinedness, those which provide an arbitrary, and usually unknown, denotation for each undefined term, and those that deal with undefined terms at the level of predicates. We have reviewed examples of all three approaches above, and shown that bunch theory provides another approach which is completely different as {\em each term has a non-arbitrary denotation}.  However, terms may be non-deterministic, which allows them to be either over-defined, to the point of non-existence, or under-defined in the case of ambiguity.
Whereas Jones et al advocate a logic adapted to undefinedness, bunch theory provides a theory of values which avoids undefinedness. In particular the use of $null$ and the improper bunch $⊥$ allows us to reason about the non-existent in a mathematically consistent way.  The main adjustment required from the user of the theory is the loss of some laws, for example $A \, < \, B \, ∨ \, A \, ≥ \, B$ which can be false when we are dealing with non-deterministic values, e.g. we have $¬(1,3 \, < \, 2)$ but also $¬(1,3 \, ≥ \, 2)$. But of course, all such laws are obeyed when we are dealing only with elements.

\subsection{Bunch theory and programming language semantics}

Within the semantics of a programming language we can use bunches to model non-determinism, and to express the results of a backtracking search carried out by a reversible computation.

Our agenda in this section will first of all be to see how 
non-deterministic or over deterministic values can be accommodated within the weakest pre-condition (WP) semantics used in the B-Method. We will then go on to derive, from predicate transformer semantics, an alternative “prospective value” (PV) semantics, equivalent in expressive power over the semantic components of B, but which also enables us to extend the expression language of B in an interesting way to express the results of a backtracking search, to express preferential choice, and to incorporate probabilistic choice along with non-deterministic choice.

Before proceeding, we review some bunch theory results needed for the development of these theories.

A bunch is said to be atomic if it cannot be formed by the union of two non-null bunches which are both different from itself. Thus $z$ is atomic if $z \, = \, A,B \, ⇒ \, z:A \, ∨ \, z:B$.

Elements are atomic; the null  bunch is atomic; the improper bunch is atomic. For the theory we present here we take the design decision that no other bunches are atomic. To achieve this we make $⊥$ absorptive in most contexts. E.g. $a \, \mapsto ⊥ \, = \, ⊥ \, $ and $⊥ \, \mapsto  \, b \, = \, ⊥ \, $. Also  $⊥ \, \setminus A \, = \, ⊥ \, $. Indeed, although $A:⊥ \, $ for any bunch $A$, if $A \, ≠ \, null$ our formalism gives us {\em no way} to express the elements that are part of $⊥ \, $ but not part of $A$. 

For any atomic bunch $z$ we have the bunch comprehension property:

\[z \, : \, ∮ \, x \, \bullet  \, E \,  \, ⇔ \,  \, (∃ \, x \, \bullet  \, z:E) \, \]

Recalling that the bunch comprehension $∮ \, x \, \bullet  \, E$ is the bunch formed from all values of $E$ as $x$ ranges over its type, the bunch comprehension property tells us that if $z$ is a sub-bunch of the bunch comprehension, there must be some value of $x$ for which $z:E$. This is intuitively obvious for elements; where $z$ is the improper bunch it means that there is some $x$ which yields the improper bunch {\em all at once}. 

Atomicity also relates to bunch inclusion, We have $A:B$ if, for an arbitrary atomic bunch $z$, $z:A \, ⇒ \, z:B$.  Bunches $A$ and $B$ are equal if $A:B \, ∧ \, B:A$. i.e. if $z:A \, ⇔ \, z:B$. 

We note that $⊥ \, : \, A$ is equivalent to $⊥ \, = \, A$, since $A \, : \, ⊥ \, $ is true for any bunch $A$.

Atomicity is mainly important because it allows us to write proofs without taking the improper bunch as a special case.

\subsection{Programming language syntax}

Here we give indicative abstract  syntax for our guarded  command language. In the  following $S$ and $T$ are commands, $g$ is a guard (i.e. a simple predicate which may be evaluated  during the execution of  a program), $P$ is a predicate, $x,y,u,v$ are a variables or comma separated variable lists, $E$ is an expression or expression list. 

The purpose of an abstract syntax is to characterise the possible parse trees of a language. This allows us to give a description that omits details of precedence and associativity (these details are provided in appendix \ref{precedence}). Our descriptions are of syntactic forms. For example when we say $S⊓T$ is an operation, this is to be taken as shorthand for “ if $S$ and $T$ are the parse trees of operations, then $S⊓T$ is also the parse tree of an operation,  having $⊓$ as its root node, $S$ as its left sub-tree, and $T$ as its right sub-tree”.

With this  in mind we can list the syntactic forms of our commands.

$Skip$ and $x:=E$ (basic commands).

$P \, | \, S$ (pre-conditioned command),  $S \, \asemi  \, T$ (sequence), $S⊓T$ (choice), $S\pref T$ (preferential choice) $S \, ⊕_p \, T$ (probabilistic choice) $g⟹S$ (guarded command).

$\IF \, g \, \THEN \, S \, \END$

$\IF \, g \, \THEN \, S \, \ELSE \, T \, \END$

$\WHILE \, g \, \DO \, S \, \END$.

The syntax for procedure definition is

$y←Op(x) \, \defs  \, P \, | \, S$

and for procedure call:

$v←Op(u)$

\subsection{Programming language semantics}

The B-Method \cite{JRA96} is based on weakest pre-condition semantics. We write $[S]Q$ for the pre-condition necessary before execution of $S$ to ensure that the post-condition $Q$ will exist after $S$ is executed. 

We will give the rules for applying this semantics over the various abstract syntax forms of our language. These will generally follow those given in \cite{JRA96} with the exception of  unpounded non-deterministic assignment, where the  presence of  bunches will make the classical rule redundant.

We will also discuss an alternative and more powerful semantics of “prospective values”, in which we write $S\diamond E$ for the value expression $E$ would take were $S$ to be executed. In this approach we also allow $S\diamond E$ as an expression, which will give us a  very  eloquent expression language that will  be illustrated in a  case study.

\subsubsection{Weakest pre-condition and conjugate weakest pre-condition  semantics}

We write $[S]Q$ for the condition that must hold before operation $S$ in order to ensure that predicate $Q$ will hold afterwards. For the assignment $x:=F$ this condition, as given in the B-Book\cite{JRA96}, is as follows:

\[[x:=F]Q \,  \, \eq \,  \, Q[F/x] \,  \, \]

Now if $F$ can be a bunch of values, each of which represents some value which $F$ might take, we require, in order to guarantee $Q$ after $S$, that $Q$ would hold for each and all of the values comprising $F$, so our law becomes:

\[[x:=F]Q \,  \, \eq \,  \, ∀ \, x' \, \bullet  \, x':F \, ⇒ \, Q[x'/x] \, \]

The B-Method has a non-deterministic assignment statement, but introducing bunches renders this redundant as we can now use a standard assignment to express unbounded non-determinism. Incorporating the new rule for non-deterministic assignment given above, and omitting a now redundant rule for unbounded non-determinism, our WP rules are as follows:

$\begin{array}{lll} \text{Name} \, & \, \text{Rule} \, & \,  \\[5pt]

\text{skip} \, & \, [skip]Q \, \eq \, Q \, & \\
\text{assignment} \, & \, [x:=F]Q \, \eq \, ∀ \, x \, \bullet  \, x:F \, ⇒ \, Q \\
\text{pre-condition} \, & \, [P|S]Q \, \eq \,  \, P \, ∧ \, [S]Q \, & \\
\text{guard} \, & \, [P \, \Guard S]Q \, \eq \, P \, ⇒ \, [S]Q \, & \,  \\
\text{bounded choice} \,  \, & \,  \, [S \, ⊓ \, T]Q \, \eq \, [S]Q \, ∧ \, [T]Q \, & \\
\text{seq comp} \, & \, [S \, \asemi  \, T]Q \,  \, \eq \,  \, [S][T]Q \, & \\
\end{array}$

We have a dual set of rules for the conjugate weakest pre-condition (CWP), $⟨S⟩Q$, the condition that $Q$ {\em might} be true after $S$. We obtain the CWP of $S$ from the WP of $S$ according to:

\[⟨S⟩Q \,  \, \eq \,  \, ¬[S]¬Q \,  \, \]

where the double negation corresponds to that seen in De Morgan's laws, or more generally to that seen in changing from universal to existential quantification. Here is the derivation of the CWP law for bounded choice:

$\begin{array}{l} ⟨S \, ⊓ \, T⟩Q \,  \, \eq \,  \, \text{“defn of CWP”} \\
¬[S⊓T]¬Q \,  \, \eq \, \text{“WP law for bounded choice”} \\
¬([S]¬Q \, ∧ \, [T]¬Q) \, \eq \,  \, \text{“logic - de Morgan”} \\
¬[S]¬Q \, ∨ \, ¬[T]¬Q \,  \, \eq \,  \, \text{“defn of CWP”} \\
⟨S⟩Q \, ∨ \, ⟨T⟩Q \\
\end{array}$

\subsubsection{Prospective value semantics}

If $S$ is an operation, we write $S \, \diamond  \, E$ for the value expression $E$ would take were program $S$ to be executed. We call $S \, \diamond  \, E$ the {\em prospective value} of $E$ following $S$. For example $x:=2 \, \diamond  \, x+10 \,  \, \eq \,  \, 12$. Where $S$ is non-deterministic, this will be reflected, for suitable $E$, in $S \, \diamond  \, E$ being a non-elemental bunch.
For example $x:=1 \, ⊓ \, x:=2 \, \diamond  \, x+10 \,  \, \eq \,  \, 11,12$ 

\subsubsection{A basic law linking prospective values and CWP}

\label{basiclaw}
We now give an intuitive law that links prospective values with CWP.  For any atomic bunch $z$ where $z$ is not free in $E$ (which  we write as $z \, \setminus E$) we have: \footnote{See \cite{arXivBunches_v2} appendix B for a proof of this law from an explicit definition of $S\diamond E$.}

\[z \, : \, (S \, \diamond  \, E) \,  \,  \, \eq \,  \, ⟨S⟩z:E \,  \,  \,  \,  \, \text{where} \, z \, \setminus E \,  \,  \,  \, \text{“The basic law”}\]

\subsubsection{Derivation of the rules for $S \, \diamond  \, E$}

We can use the basic law to derive the following rules.  \label{primitives}

$\begin{array}{lll} \text{Name} \, & \, \text{Rule} \, & \,  \\[5pt]

\text{skip} \, & \, skip \, \diamond  \, E \,  \, \eq \,  \, E \, & \\
\text{assignment} \, & \, x:=F \, \diamond  \, E \,  \, \eq \, (λx.E)(F) \\
\text{pre-condition} \, & \, P \, | \, S \, \diamond  \, E \,  \, \eq \,  \, P \, \bpre  \, (S \, \diamond  \, E) \, & \\
\text{guard} \, & \, P \, \Guard S \, \diamond  \, E \,  \, \eq \,  \, P \, \bguard  \, (S \, \diamond  \, E) \, & \,  \\
\text{bounded choice} \,  \, & \,  \, S \, ⊓ \, T \, \diamond  \, E \,  \, \eq \,  \, (S \, \diamond  \, E) \, , \, (T \, \diamond  \, E) \,  \\
\text{seq comp} \, & \, S \, \asemi  \, T \, \diamond  \, E \,  \, \eq \,  \, S \, \diamond  \, T \, \diamond  \, E& \\
\end{array}$

The precedence of $\diamond $ is below that of programming connectives, whose precedence, in descending order, is $:= \, , \, \Guard , \, ⊓ \, , \, \asemi  \, , \, |$. We follow Hehner in introducing a large equals $\eq$, which is a low precedence symbol used in equational reasoning. When written between expression or predicates it indicates that they are equivalent. 

Our derivations are based on the proposition that for any bunches $A$ and $B$, and for any atomic bunch $z$ we have:

$A \, = \, B \,  \, ⇔ \,  \, (z:A \, ⇔ \, z:B)$.

We provide example derivations for assignment and sequential composition.\footnote{A full set of derivations is provided in \cite{arXivBunches_v2}.}

For assignment and unbounded choice we have:

$\begin{array}{l} z \, : \, (x:=F \, \diamond  \, E) \,  \, \eq \,  \, \text{“basic law, see section \ref{basiclaw}”} \\
⟨x:=F⟩z:E \,  \, \eq \,  \, \text{“CWP rule} \, ⟨x:=F⟩Q \, ⇔ \, (∃x\bullet x:F \, ∧ \, Q) \, \text{”} \\
∃x \, \bullet  \, x:F \, ∧ \, z:E \,  \, \eq \,  \, \text{“property of bunch guard”} \\
∃x \, \bullet  \, z \, : \, x:F \, \bguard  \, E \,  \, \eq \,  \, \text{“property of bunch comprehension”} \\
z \, : \, ∮ \, x \, \bullet  \, x:F \, \bguard  \, E \,  \, \text{“and thus”} \\
x:=F \, \diamond  \, E \,  \, \eq \,  \, ∮ \, x \, \bullet  \, x:F \, \bguard  \, E \,  \\
\end{array}$

And for sequential composition:

$\begin{array}{l} z \, : \, (S \, \asemi  \, T \, \diamond  \, E) \,  \, \eq \,  \, \text{“basic law”} \\
⟨S \, \asemi  \, T⟩z:E \,  \,  \, \eq \,  \, \text{“CWP rule} \, ⟨S\asemi T⟩Q \, ⇔ \, ⟨S⟩⟨T⟩Q \, \text{”} \\
⟨S⟩⟨T⟩z:E \,  \,  \,  \, \eq \,  \, \text{”basic law“} \\
⟨S⟩ \, z \, : \, (T \, \diamond  \, E) \,  \, \eq \,  \, \text{“basic law and right assoc of} \, \diamond  \, \text{”} \\
z \, : \, (S \, \diamond  \, T \, \diamond  \, E) \,  \, \text{“and thus”} \\
S \, \asemi  \, T \, \diamond  \, E \,  \,  \, \eq \,  \, S \, \diamond  \, T \, \diamond  \, E \,  \\
\end{array}$

\subsection{Bunches, homogeneity and the avoidance of complexity}

\label{simplicity}

The contribution of bunch theory to the simplicity of our prospective-value semantics can be seen in the simplicity of the rules for $skip$ and for sequential composition compared to the rules that would be required were we to use classical set theory. In the latter case, to accommodate non-determinism, we require a set to  express the possible values taken by expression $E$ following $S$. Both forms of the rules are given in the following table.

$\begin{array}{lll}  \\
\text{Name} \, & \, \text{Bunch Rule} \, & \, \text{Set Rule} \\
\hline \, \vspace{-10pt} \\
\text{Skip} \, & \, skip \, \diamond  \, E \, \eq \, E \,  \, & \,  \, skip \, \diamond  \, E \, \eq \, \{ \, E \, \} \\
\text{Seq comp } \, & \, S \, \asemi  \, T \, \diamond  \, E \, \eq \, S \, \diamond  \, T \, \diamond  \, E \,  \, & \,  \, S \, \asemi  \, T \, \diamond  \, E \, \eq \, ⋃(S\diamond T\diamond E) \\
\end{array}$

The set rule for $skip$ illustrates a loss of homogeneity, which has a corresponding effect on the set rule for sequential composition. We can illustrate the added complexity  of  the set form by analysing $skip\asemi skip \, \diamond  \, E$ in both cases. Using the bunch rule we  have:

$\begin{array}{l} skip\asemi skip \, \diamond  \, E \,  \,  \, \eq \,  \, skip \, \diamond  \, skip \, \diamond  \, E \,  \, \eq \,  \, skip \, \diamond  \, E \,  \, \eq \, E \,  \,  \,  \,  \,  \\
\end{array}$

and for the set rule

$\begin{array}{l} skip\asemi skip \, \diamond  \, E \, \eq \,  \, ⋃ \, skip \, \diamond  \, skip \, \diamond  \, E \,  \, \eq \, ⋃ \, skip \, \diamond  \, \{E\} \,  \, \eq \,  \, ⋃ \, \{\{E\}\} \\
\eq \, \{E\} \\
\end{array}$

\subsection{Selection and iteration}

The attentive reader will have noted that the semantic components do not include selection and iteration constructs. These are handled by means of choice and guard, as in the B-Method \cite{JRA96}.

The $\IF$ construct is defined as:

\[\IF \, g \, \THEN \, S \, \ELSE \, T \, \END \,  \, \defs  \,  \, g \, \Guard S \,  \, ⊓ \,  \,  \, ¬g \, \Guard T \, \]

Let's see how this works when evaluating the prospective value
of expression $E$ after an $\IF$ construct:

$\begin{array}{l}  \,  \, \IF \, g \, \THEN \, S \, \ELSE \, T \, \END \, \diamond  \, E \, \eq \, `` \text{by defn of }\IF\text{”} \\
 \,  \, g \, \Guard S \,  \, ⊓ \,  \,  \, ¬g \, \Guard T \,  \, \diamond  \, E\eq \,  \, \q{by rule for bounded choice} \\
 \,  \, (g \, \Guard S \, \diamond  \, E) \, , \, (¬g \, \Guard T \, \diamond  \, E) \,  \, \eq \,  \, \q{by rule for guard} \\
 \,  \, g \, \bguard  \, (S \, \diamond  \, E) \, , \, ¬g \, \bguard  \, (T \, \diamond  \, E) \\
\end{array}$

When $g$ is true, the properties of the bunch guard tell us that this reduces to $(S \, \diamond  \, E) \, , \, null$, i.e. to $S \, \diamond  \, E$. In this case the $\ELSE$ clause contributes nothing to the result, and our ability to express this nothing by means of the empty bunch $null$ is essential to our formulation. Similarly, when $g$ is false the construct reduces to $null \, , \, (T \, \diamond  \, E)$, i.e. to $T \, \diamond  \, E$. 

We have an iteration construct

$\WHILE \, g \, \DO \,  \, S \, \END$, which  we will refer to  as $W$.

We define $W$ recursively by unwrapping the first iteration of  the  loop:

$W \, \eq \, \IF \, g \, \THEN \, S \, \asemi  \, W \, \ELSE \, skip \, \END$

Giving us a recursive equation in programs. That such an equation has a solution is shown in the B-Book. We give a bunch theoretic treatment in appendix \ref{Fixpoint}.

\subsection{Provisional choice and backtracking}

The semantics of choice, which we have used to incorporate demonic non-determinism into our language, can be used to describe provisional choice, as used in a backtracking application. This idea has a long history. As far back as 1967, in his paper ``Non-deterministic Algorithms'' \cite{RWF67}, Floyd talked of ``programs governed in part .. by final causes for the sake of which their effects are carried out''. In ``The Specification Statement''\cite{CM88} Carroll Morgan mentions the possibility as follows: ``Ordinarily we limit the syntax of our programming language so that miracles cannot be written in it. If we relax this restriction, allowing naked guarded commands, then operational reasoning suggests a backtracking implementation''. Backtracking constructs are also discussed by Nelson \cite{GN89}, and Norvel and Hehner \cite{NH92} \cite{ECRH93}.

We have explored the possibility of exploiting a prospective value semantics to represent backtracking, and in particular the implementation of backtracking via reversible computing \cite{SZL06,SZD10,FZThesis}. Operationally, a naked guarded command $g \, \Guard S$ will provoke reversal of computation if $g$ is false, and will execute as $S$ if $g$ is true. Also, choice from a set $x \, :∈ \, E$ will provoke reversal of computation if the expression $E$ denotes an empty  set. Reverse computation undoes any changes made during  forward execution, and continues until a previous provisional choice is reached, at which point forward execution will recommence if a previously untried choice is available. Should there be no untried choices available and the program has been executed from a terminal, it returns to the user but with a different prompt, e.g. “ko” rather than “ok”, to indicate execution was infeasible.

The most novel feature of our proposed language is  the incorporation, within our sub-language of expressions, of  terms of the  form $\{ \, S \, \diamond  \, E \, \}$. These greatly increase the expressive power of our language, as illustrated  in the following case  study. \footnote{We have developed an implementation platform for this approach, in the form of a reversible virtual machine \cite{RVM2010}.} 

\subsection{Case study}

In the Knight's Tour problem, our task is  to  construct a series of moves for  a knight on a chess board so that after 64 moves the knight has  completed a tour which visits every square  on the board, and has returned to the original square. With a  backtracking language this can be achieved by programming a loop which at every  iteration makes  a provisional move to an unvisited square, backtracks if no  such square is available, and  terminates when it returns  to the starting  square. Eventually such an approach will find a  tour, but its efficiency can be greatly increased by  employing the “Wansdorf Heuristic” which requires us to choose moves that take as to the most tightly constrained squares, on the grounds that these are the  squares most at risk of becoming dead ends. The implementation of this  heuristic will comprise our case study. We assume a global integer program variable $square$ which holds the current position of the knight, and an operation $Move$ which will make a provisional choice of a move to a new, previously unvisited square. The set of all possible squares the knight can move to is given by the term $\{ \, Move \, \diamond  \, square \, \}$. The size of this set tells us  how constrained the current square is. To implement our heuristic we construct a relation between the possible squares available for the knight  to  move  to, and how constrained each of these squares is, and from this we obtain the set of most tightly constrained squares available for our next move.

$\begin{array}{l}  \,  \, \t \,  \,  \, \bs\text{Lines starting with }\bs\text{ are comments} \\
 \,  \, \t \,  \,  \, \bs \, available\text{ is the set of squares we can move to} \\
 \,  \, \t \,  \, available \, := \, \{move\diamond square\} \, \asemi  \,  \,  \, available \, > \, 0 \, \Guard skip \, \asemi  \\
 \,  \, \t \,  \, \bs\text{We create a function mapping available squares to the number of} \,  \\
 \,  \, \t \,  \, \bs\text{onward moves that  can be made from each} \\
 \,  \, \t \,  \, constraints \, := \, \{ \, square \, :∈ \, available \, \diamond  \, square \, \mapsto  \, \card(\{move \, \diamond  \, square\}) \, \} \, \asemi  \\
 \,  \, \t \,  \, \bs \, m \, \text{ is the least number of onward moves that can be made} \\
 \,  \, \t \,  \, \bs\text{from an available square} \\
 \,  \, \t \,  \, m \, := \, minset(range(constraints)) \, \asemi  \\
 \,  \, \t \,  \, \bs\text{we use domain restriction to limit our function to squares having $m$} \\
 \,  \, \t \,  \, \bs\text{onward moves, then take its domain.} \\
 \,  \, \t \,  \, \bs \, tight \, \text{ is the set of squares that only  have }m\text{ onward moves} \\
 \,  \, \t \,  \, tight \, = \, domain(constraints \, \rres  \, \{m\}) \, \asemi  \,  \,  \\
\end{array}$

In the computation of a term of the form $\{S \, \diamond  \, E\}$, each time a way is found to complete the execution of $S$, the value of $E$ in the resulting after-state is calculated and added to the set of results being constructed. Execution is then reversed, so that all possible choices are explored. The topology of this execution is in the form of a tree, and ends when backwards execution finds no more unexplored choices. The value of the term $\{S \, \diamond  \, E\}$ is then known, and forward execution can recommence {\em without the evaluation of term} $\{S \, \diamond  \, E\}$ having caused any change in the program state (or, operationally speaking, any enduring change, since any changes are restored during reverse execution).

Thus we have a simple semantics which allows us to express non-determinism and demonic choice along with provisional choice and backtracking within the same development framework. However, in a term of the form $\{S \, \diamond  \, E\}$ any choice within $S$ must be considered as provisional choice rather than  implementer's choice, and must not be subject to refinement.

\subsection{Preference as a refinement of provisional choice }

In G Nelson's generalisation of Dijkstra's WP calculus \cite{GN89}, we find a biased choice operator $S \, ⊞ \, T$. In a backtracking scenario, $S$ would be chosen in preference to $T$, but if $S$ is unable to produce a solution, $T$ will be tried instead. Unfortunately, Nelson's construct is limited by the fact that it is only $S$ which has to be able to find a solution. We would like to extend this idea to take account not only of $S$, but also of the rest of the program which follows. Nelson's biased choice is defined using the {\em feasibility} of $S$. We will first examine the idea of feasibility, and then extend it  {\em to take account of the effect of any further computation,} which we refer to as the continuation. To achieve this we will use a construct which cannot be formulated in classical WP semantics, but which is made possible by our use of PV semantics and bunches. 

Following Abrial we write the feasibility of $S$ as $fis(S)$ and recall that a program is feasible if it cannot establish false, i.e.  $fis(S) \, \defs  \, ¬[S]false$.

We also recall a monotonicity condition of the WP calculus, that if predicate $Q'$ is weaker than $Q$, i.e. if $Q \, ⇒ \, Q'$, a program able to establish $Q$ will also be able to establish $Q'$. Thus, since $false \, ⇒ \, Q$ for any predicate $Q$, a program able to establish the post condition $false$ could establish absolutely anything. The truths established in this way, often known as “miracles” \cite{MRG88}, are vacuous, and depend on the computation having no after states. Thus they may be compared to the true result obtained from universal quantification over an empty set. These {\em vacuous} truths play an essential  role in our semantics. For example, in the decomposition, first proposed by Hehner\footnote{As recounted in \cite{HEHNER06}, this idea was first made public in a technical report because the reviewers of the paper in which it was proposed required it to be removed as a condition of acceptance.}, of a conditional $\IF \, g \, \THEN \, S \, \ELSE \, T \, \END$ into the guarded choice $g \, \Guard S \, ⊓ \, ¬g \, \Guard T$, we use a vacuous truth to express the contribution of the choice {\em which is not taken}. 

In a backtracking interpretation, a program is infeasible if it is always forced to backtrack, and accordingly has no after-states. Using this idea, Nelson's biased choice can be defined as:

\[S \, ⊞ \, T \,  \, \defs  \,  \, S \, ⊓ \, ¬fis(S) \, \Guard T \, \]

And we can clearly see that it is the feasibility of $S$ alone that guides the choice. We want to extend the idea of feasibility so it applies to the rest of the program, but first let us see how the predicate $¬fis(S)$ might be expressed in PV semantics. Starting from our CWP definition $fis(S) \, \defs  \, ⟨S⟩true$, we have:

$\begin{array}{l} ¬fis(S) \,  \, \eq \,  \, \text{“by defn of }fis(S)\text{”} \,  \\
¬¬[S]false \,  \, \eq \,  \,  \, \text{“expressing in CWP”} \,  \,  \\
¬⟨S⟩true \,  \, \eq \,  \, \text{“since }z:⊥ \,  \, \eq \,  \, true\text{”} \,  \,  \\
¬⟨S⟩z:⊥ \,  \,  \, \eq \,  \, \text{“basic law”} \,  \,  \\
¬ \, z \, : \, (S\diamond ⊥) \,  \,  \, \eq \,  \, \text{“since }z\text{ is arbitrary”} \,  \,  \\
(S\diamond ⊥ \, ) \, : \, null \,  \\
\end{array}$

We would like to design a similar form of choice, which we will write as $\drangle $, but rather than looking just at the feasibility of $S$, we would like to consider the feasibility of $S$ along with any continuation of the program after the choice. Thus for the program  $S \, \drangle  \, T \, \asemi  \, U$,  (where choice “$ \, \drangle  \, $” binds more tightly than operation sequence “$ \, \asemi  \, $”) we would be interested in the feasibility of $S\asemi U$, and it would be the infeasibility of $S\asemi U$ that would force the choice of $T$.  We propose the following definition  of preferential choice

\[S \, \drangle  \, T \, \diamond  \, E \,  \, \eq \,  \, (S\diamond E) \, , \, (S\diamond E)=null \, \bguard  \, (T\diamond E) \, \]

This executes $S$ in preference to $T$, but if execution of $S$ or its continuation results in backtracking to this point, $T$ is then executed. A simple example will illustrate how bunch theory allows the expression $E$ to represent the feasibility of a continuation. We look at a program that would prefer the assignment $x:=1$ to $x:=2$, but is forced by a {\em future} guard to make the alternative choice.

\[ x:=1 \, \drangle  \, x:=2 \,  \, \asemi  \,  \, x=2\Guard skip \,  \, \diamond  \,  \, x \,  \,  \,  \,  \,  \, \eq \,  \, \text{“PV rule for seq comp”} \\
x:=1 \, \drangle  \, x:=2 \,  \, \diamond  \,  \, x=2\Guard skip \, \diamond  \,  \, x \,  \,  \,  \,  \,  \, \eq \,  \, \text{“PV rule for guard”} \\
x:=1 \, \drangle  \, x:=2 \,  \, \diamond  \,  \, x=2 \, \bguard  \, x \,  \,  \, \eq \,  \, \text{“preferential choice”} \\
(x:=1 \, \diamond  \, x=2 \, \bguard  \, x), \,  \\
(x:=1 \, \diamond  \, x=2 \, \bguard  \, x)=null \, \bguard  \, (x:=2 \, \diamond  \,  \, x=2 \, \bguard  \, x) \,  \,  \\
\eq \,  \, \text{“PV rule for assignment x2”} \\
(1=2\bguard  \, 1), \, ((1=2\bguard  \, 1)=null) \,  \, \bguard  \, (2=2\bguard  \, 2) \,  \,  \,  \\
\eq \, \text{“evaluating guards”} \\
null \, , \, (null=null) \, \bguard  \, (true \, \bguard  \, 2) \,  \,  \, \eq \,  \, \text{“further evaluation”} \\
2\]

In this example the message concerning the future evolution of the computation after the preferential choice is contained in the expression $x=2 \, \bguard  \, x$. This has the value $null$ (i.e. expresses non-existence) unless $x=2$, and in this way conveys that future execution must backtrack unless $x=2$. 

We have seen that our attempt to capture preference in WP led us only to biased choice; we were unable to capture preference in WP. However, preference occurs naturally in implementations, since these are intrinsically deterministic. We were able to capture preference in our PV semantics because in the term $S \, \diamond  \, E$ the expression $E$ is able to represent the total future behaviour of the code following $S$ (its continuation). When $E=null$ this continuation is infeasible.

Preferential choice is easily shown to be a refinement of provisional choice. The condition that program $S$ is refined by program $T$, when both programs have a state variable or variable list $x$, is expressed  in terms of PV semantics as\footnote{For a full discussion see appendix  C of \cite{arXivBunches_v2}.}

\( \, S \, ⊑ \, T \,  \, \defs  \,  \, (T \, \diamond  \, E) \,  \, : \,  \, (S \, \diamond  \, E) \, \)

We can argue as follows:

$\begin{array}{l} S \, \drangle  \, T \, \diamond  \, E \,  \, \eq \, \text{“defn of preferential choice”} \\
(S\diamond E),(S\diamond E)=null \, \bguard  \, (T\diamond E) \,  \, : \,  \, \text{“which  is a sub-bunch of”} \,  \\
\text{“since }g \, \bguard  \, β \, : \, β \, \text{ and } \, β' \, : \, β \, ⇒ \, α,β' \, : \, α,β \, \text{”} \\
(S\diamond E),(T\diamond E) \,  \, \eq \, \text{“PV defn of non-deterministic choice”} \\
S \, ⊓ \, T \, \diamond  \, E \,  \\
\end{array}$

\subsection{Using bunch properties in the formulation of probabilistic choice.} 

In classical probability theory, if $X$ is a random variable that takes value $x_1{}$ with probability $p$ and value $x_2{}$ with probability $1-p$, then the “expected value” of $X$ is given by the weighted addition $\Exp \, (X) \, = \, x_1{} \,  \, _p\!\!+ \, x_2{}$

where $x_1{} \,  \, _p\!\!+ \,  \, x_2{} \,  \, = \,  \, p*x_1{} \, + \, (1-p)*x_2{}$

In \cite{SPECIALPROB} we formulate a probabilistic choice between programs, where for feasible continuations, $S \, _p\!⊕ \, T$ will choose execution of $S$ with probability $p$ and $T$ with probability $1-p$. We define the expectation of some expression $X$ (of suitable type) after executing such a choice as a weighted addition. For $0<p<1$ this expectation\footnote{In \cite{SPECIALPROB} we use the notation $\Exp(S\diamond X)$ for the expected value of $X$ after executing $S$. That notation resembles the $E(X)$ of probability theory, but is not compositional, so we change it here to $S \, \diamond _E \, X$.} is given by:

\[S \, _p\!⊕ \, T \, \diamond _E \, X \, \defs  \, (S \, \diamond _E \, X) \,  \, _p\!\!+ \,  \, (T \, \diamond _E \, X) \, \]

Where $S$ and/or $T$ involve non-determinism their expectations will be a bunch of values, and the bunch lifting of $_p\!+$ is all that is needed to integrate probabilistic and non-deterministic choice.

However, that is not quite enough to describe the backtracking behaviour of our reversible machine. If a probabilistic choice leads to infeasibility, then execution will reverse to the point of probabilistic choice and take the other choice. We incorporate this into our weighted choice by expressing a weighted choice as a bunch of three terms. We are using the same technique we previously showed for preferential choice.

\[X_1{} \, _p\!\!+ \, X_2{} \, \defs  \, X_1{} \, = \, null \, \bguard  \, X_2{} \, , \, X_2{} \, = \, null \, \bguard  \,  \, X_1{} \, , \,  \\
 \,  \,  \,  \,  \,  \,  \,  \,  \,  \,  \, p*X_1{}+(1-p)*X_2{} \, \]

Where $X_1{}$ and $X_2{}$ are non-null, this corresponds to the probabilistic choice not leading  to infeasibility. The first two terms then equate to $null$ and do not contribute to the result, which is given by the third term. If either $X_1{}$ or $X_2{}$ is $null$, then, by the absorptive properties of $null$, the third term will be $null$  and the result will be given by the first two terms, at most one of which will be non-null.

Finally we need to consider what our expectation should be if a probabilistic choice could lead to an aborting computation. Our approach is that any possibility of abort is counted as abort. To achieve this needs no further modification of our weighted addition. If either $X_1{}$ or $X_2{}$ describes an abortive continuation, its value will be $⊥ \, $. The term $p*X_1{}+(1-p)*X_2{}$ will then be $⊥ \, $ (by the absorptive power of $⊥ \, $), and the whole weighted addition will equate to $⊥ \, $. 

Through use of bunches we have been able to capture interactions between probabilistic choice, non-deterministic choice, feasibility and aborting behaviour {\em in the world of values.}

\section{Extending axiomatic set theory to incorporate bunch theory}

There are two aspects to extending set theory to incorporate bunch theory.

The first is to add new axioms which are specific to bunches, and make a slight adjustment to one of our existing axioms.

The second is to weaken classical logic FOPL to FOPLN, which is necessary because we can no longer assume that every expression denotes an element.

\subsection{Our axioms} 

Hehner gives an axiomatisation in which bunch theory is presented as an independent, stand alone theory, to which the axioms of set theory may subsequently be added. Here we take a different approach in which we add axioms to set theory to capture the idea of a bunch as the contents of a set. We choose the formalisation of set theory given by Abrial in the B-Book\cite{JRA96}\footnote{We conjecture that the same approach can be used with any axiomatisation of set theory.}.

We must express one of Abrial's axioms differently for it to be meaningful in the language of bunch theory. In the B-Book\cite{JRA96} Abrial writes the axiom of choice as: $∃x \, \bullet  \, x \, ∈ \, s \, ⇒ \, choice(s) \, ∈ \, s$ where $x∖s$ \footnote{Notation: we use $x∖E$ to indicate that identifier $x$ must not occur free in expression $E$}. Since $null∈s$ vacuousy in bunch theory, once bunches are introduced Abrial's form of  the axiom allows $choice(s) \, = \, null$. We use an equivalent form that forces $choice(s)$ to be an element. To prove the  equivalence of the two forms of this axiom, abstract them as $ \, a \, ⇒ \, b$ and $a \, ⇒ \, c$, note that $b \, ⇔ \, c$ by the one point rule,\footnote{$(∃ \, x \, \bullet  \,  \, x=E \, ∧ \, P \, ) \,  \, ⇔ \,  \, P[E/x] \,  \,  \,  \,  \, x∖E$} then note that by  propositional logic:

$(b \, ⇔ \, c) \, ⇒ \, (a⇒b \, ⇔ \, a⇒c)$ 

We now state our axioms:

$\begin{array}{lll} \text{Name} \, & \, \text{Rule} \, & \, \text{Condition} \\
\text{Ordered pair} \, & \, E\mapsto F \, ∈ \, (s \, × \, t) \, ⇔ \, E \, ∈ \, s \, ∧ \, F \, ∈ \, t \\
\text{Powerset} \,  \,  \,  \,  \, & \, s \, ∈ \, \pow (t) \,  \, ⇔ \,  \, (∀ \, x \, \bullet  \, x \, ∈ \, s \, ⇒ \, x \, ∈ \, t) \, & \, x∖(s,t) \\
\text{Comprehension} \, & \, E \, ∈ \, \{x|x \, ∈ \, s \, ∧ \, P\} \, ⇔ \, E \, ∈ \, s \,  \, ∧ \,  \, P[E/x] \, & \, x∖s \\
\text{Set equality} \, & \,  \, (∀ \, x \, \bullet  \, x \, ∈ \, s \, ⇔ \, x \, ∈ \, t) \, ⇒ \, s=t \, & \, x∖(s,t) \\
\text{Choice} \, & \, (∃x \, \bullet  \, x \, ∈ \, s) \, ⇒ \, (∃x \, \bullet  \, x \, ∈ \, s \, ∧ \, x \, = \, choice(s)) \,  \, & \, x∖s \\
\text{BIG} \,  \,  \,  \, & \, infinite(BIG) \,  \\
\end{array}$

We follow the general  mathematical convention, also used by Abrial, by which variables may  be implicitly  universally quantified at the top level.\footnote{By their nature, axioms do not include  any free variables.} This applies, for example, to the variables $E,F,s$ and $t$ in the ordered pair axiom. An added detail in bunch  theory is that, since quantification takes place over elements, such impicitly quantified variables will always represent elements.

We add two axioms to characterise the meaning of infinite, a detail which is
omitted from the B Book.

$\begin{array}{lll} \text{Infinity 1} \, & \, infinite(s) \, ⇒ \, ∃ \, x \, \bullet  \, x \, ∈ \, s \\
\text{Infinity 2} \, & \, infinite(s) \, ⇔ \, infinite(\{x \, | \, x \, ∈ \, s \, ∧ \, x \, ≠ \, choice(s)\}) \\
\end{array}$

We now give the additional axioms which enable us to incorporate bunch theory. We keep the number of axioms to a minimum and we rely on post-hoc definitions to introduce many of the notations associated with bunches, including bunch union and intersection, bunch inclusion, the null bunch, and the lifted nature of set membership within our bunch theory. This approach facilitates validation of our model and lowers the risk of missing axioms - since what might have been candidates for axioms become properties provable from post-hoc definitions.

$\begin{array}{lll} \text{Name} \, & \, \text{Rule} \, & \, \text{Condition} \,  \\
\text{Packaging 1} \, & \, \{\unpack A\} \, = \, A \, & \, A \, \text{an element and a set} \,  \\
\text{Packaging 2} \, & \, \unpack \{E\} \, = \, E \,  \\
\text{Element 1} \, & \, \unpack A \, \text{is an element iff} \, A \, \text{is a singleton set.} \,  \\
\text{Element 2} \, & \, \{E\} \, \text{is an element} \\
\text{Guard 1} \, & \, g \,  \,  \,  \,  \, ⇒ \,  \,  \,  \,  \, g \, \bguard  \, E \,  \, = \,  \, E \\
\text{Guard 2} \, & \, ¬g \,  \,  \,  \,  \, ⇒ \,  \,  \,  \,  \, g \, \bguard  \, E \,  \, = \,  \, \unpack \{ \, \} \\
\end{array}$ 

The packaging axioms introduce the unpacking operator $\unpack $, and the use of set brackets to “package” a bunch.  They tell us that packaging and unpackaging are inverses. That is not quite enough to characterise packaging - indeed these axioms would be true if neither packaging nor unpackaging had any effect; this is ruled out by the element axioms. The first element axiom tells us that the landscape of set theory extended with bunches contains mathematical objects not present in classical set theory, namely those values obtained by unpacking non-singleton sets. The second element axiom tells us packaging creates an element - this axiom will need to be qualified when we extend our theory further with the introduction of the improper bunch. 

In the second guarded bunch axiom the term $\unpack \{ \, \}$ can be read as $null$, but we cannot write it that way as $null$ is not yet defined. 

\subsection{Weakening classical logic to  deal with undefinedness}

We need to weaken the FOPL inference rules for quantifiers to obtain those we need for FOPLN. For a more complete discussion see \cite{arXivLogic}.

The ∃-intro rule of FOPL says that if a proposition $P$ is true when $x$ is replaced by some expression $E$, then we can say that there exists some $x$ for which $P$ is true.  The ∃-intro rule for FOPL is:

$\begin{array}{l} \displaystyle\frac{HYP \, ⊢ \,  \, P[E/X]}{HYP \, ⊢ \, ∃ \, x \, \bullet  \, P} \,  \, \text{       ∃-intro} \\
\end{array}$

However, once we are dealing with bunches an expression may represent $null$, and $P[E/x]$ will then be vacuously true, and provide no evidence that some $x$ exists such that $P$ is true. Our ∃-into rule uses an additional premise  that the expression we are citing is an element.

$\begin{array}{l} \displaystyle\frac{HYP \, ⊢ \, δ(E) \, , \,  \, HYP \, ⊢ \,  \, P[E/X]}{HYP \, ⊢ \, ∃ \, x \, \bullet  \, P} \,  \, \text{       ∃-intro} \\
\end{array}$

The ∃-elim rule of FOPL is gives a generalised form of argument by cases. Suppose $x$ is not free in $Q$ (which we will write as $x∖Q$), and that according to the hypotheses under which we are reasoning we can prove  $P[α/x]⇒Q$ where $α$  is fresh, and suppose also that under the same hypotheses we can prove $∃ \, x \, \bullet  \, P$, then we can conclude $Q$. 

The ∃-elim rule for FOPL is: 

$\begin{array}{l} \displaystyle\frac{HYP \, ⊢ \, P[α/x] \, ⇒ \, Q, \,  \, HYP \, ⊢ \, ∃ \, x \, \bullet  \, P}{HYP \, ⊢ \, Q} \,  \, α \, \text{fresh,} \,  \, x∖Q \,  \,  \, \text{      ∃-elim} \\
\end{array}$

The ∃-elim rule for FOPLN provides an additional hypothesis $δ(α)$ which we can use in establishing its conclusion.

The ∃-elim rule for FOPLN is:

$\begin{array}{l} \displaystyle\frac{HYP, \, δ(α) \, ⊢ \, P[α/x] \, ⇒ \, Q, \,  \, HYP, \, δ(α) \, ⊢ \, ∃ \, x \, \bullet  \, P}{HYP \, ⊢ \, Q} \,  \, α \, \text{fresh,} \,  \, x∖Q \,  \,  \, \text{     ∃-elim} \\
\end{array}$

For a full description of FOPLN see \cite{arXivLogic}. Similar adjustments are required for the universal quantification rules, but it will be sufficient to demonstrate the soundness of its existential quantifier rules to show the soundness of FOPLN

\section{Model}

The aim of our model is to give the meaning of any expression, predicate or inference rule from our language of bunch theory underpinned by the logic FOPLN, in terms of the well established “meta-language” of set theory. 

The set theory we use for our meta-language will support “environments”. An environment is akin to a tuple, but with its components accessed by name rather than position. We use environments to associate the names of free variables in our language with the values they denote in the meta-language of our model.

If we are considering a free variable $x$ that takes a value $1,2$, we have the problem that the bunch value $1,2$ is not something that exists in the meta-language of our model. However, we can model it with the set $\{1,2\}$. Proceeding in this way we associate every value in our language will a unique and distinct denotation in our meta-language.

An environment which associates the name $x$ with  $1,2$ and the name $y$ with 3 will be written as:

\( \, \lbb  \, x \, \leadsto  \, \{1,2\}, \,  \, y \, \leadsto  \, \{3\} \, \rbb \, \)

Values of the components of an environment are obtained by projection. e.g.

\( \, \lbb  \, x \, \leadsto  \, \{1,2\}, \,  \, y \, \leadsto  \, \{3\} \, \rbb \, .x \,  \, = \,  \, \{1,2\}\)

If $E$ is an expression in our language, and $\rho $ an environment that associates values with all free variables of $E$, then $\llbracket E\rrbracket ^\nu (\rho )$ will be the value denoted by $E$ in our model when its free variables are bound to values given in the environment $\rho $. For example if $\rho  \, = \, \lbb  \, x \, \leadsto  \, \{1,2\}, \,  \, y \, \leadsto  \, \{3\} \, \rbb$ then

$\llbracket x+y\rrbracket ^\nu (\rho ) \, = \, \{4,5\}$

Because we distinguish predicates from values, our semantics for predicates differs from our semantics of values. $\llbracket P\rrbracket ^\tau (\mathcal{E} )$ represents the ``truth denotation” of predicate $P$, over the set of environments $\mathcal{E} $, its value being the subset of $\mathcal{E} $ consisting of those environments in which $P$ is true. Here are two examples in which $\rho $ is defined as above and $\mathcal{E} $ is a singleton set:

$\llbracket x \, = \, y\rrbracket ^\tau (\{ \, \rho  \, \}) \, = \, \{ \, \}$

$\llbracket x \, < \, y\rrbracket ^\tau (\{ \, \rho  \, \}) \, = \, \{ \, \rho  \, \}$

Within quantified predicates we may have to deal with variable capture, where a quantified variable has the same name as some variable from an outer scope, or we may experience an extension of an environment, where newly quantified variables are added. To deal with these situations it is useful to equip our meta-language with “binding override”, operation $⊕$. Here an example of its use: 

\( \, \lbb  \, x \, \leadsto  \, \{1,2\}, \,  \, y \, \leadsto  \, \{3\} \, \rbb \, ⊕ \,  \, \lbb  \,  \, y \, \leadsto  \, \{4,5\}, \, z \, \leadsto  \, \{2,5,7\} \, \rbb \, = \, \)

\( \,  \, \lbb  \, x \, \leadsto  \, \{1,2\}, \,  \, y \, \leadsto  \, \{4,5\}, \, z \, \leadsto  \, \{2,5,7\} \, \rbb \, \)

We provide rules to give the meaning of any expression, predicate or inference rule in our target  language in terms of its syntactic components: e.g. we would give the meaning of an expression that has the form $E \, \mapsto  \, F$ in terms of the meanings of $E$ and $F$ and the semantics of the maplet symbol. Thus our semantics is compositional and based on the denotational rules we give for each abstract syntax form in our language. 

\subsection{The rules of our model}

In the following, $c$ is an element and a constant; $x$ a variable; $E, \, F$ are general expressions; $s, \, t$ are elements and sets, $\rho $ is an environment, and $\mathcal{E} $ is a set of environments.

In the term $\llbracket E\rrbracket ^\nu (\rho )$, $E$ is an expression, and $\rho $ an environment which associates the free variables of $E$ with particular values.  $\llbracket E\rrbracket ^\nu (\rho )$ represents the “value denotation” of expression $E$ in environment $\rho $. 

Environments exist in the mathematical world of our meta language of set theory, but their formation records values taken by variables in our bunch theory. The binding $x \, \leadsto  \, \{1,2\}$ records that the variable $x$ is associated with the bunch $1,2$.

If $E$ is composed entirely of constant terms, for example $3\mapsto 4$, it will not depend on an environment, and we can write its denotation as just $\llbracket E\rrbracket $.

We present our rules a few at a time, and before each set of rules we explain the intuitions behind their formulation.  
                                                                                                                                                                                                                   
We first give the denotations for constants and variables.

The denotation of a constant is independent of its environment, for example 3 will denote $\{3\}$ in the meta language of our model. Our semantics wraps each constant term in set brackets.

\( \, \llbracket c\rrbracket ^\nu (\rho ) \, = \, \llbracket c\rrbracket  \,  \, = \, \{c\} \, \)

For variables, the value represented is given directly by the environment. No wrapping in set brackets is needed in the following rule, as this is already present within the variable's binding.

\( \, \llbracket x\rrbracket ^\nu (\rho ) \, = \, \rho .x \, \)

We next give denotations for packing and unpacking.

To obtain the denotation of $\{E\}$ we wrap the 
denotation of $E$ in brackets.

\( \, \llbracket \{E\}\rrbracket ^\nu (\rho ) \, = \, \{ \, \llbracket E\rrbracket ^\nu (\rho ) \, \} \, \)

For the denotation of unpacking we consider for the moment the unpacking of a single set $s$.  The denotation of $s$ will be a singleton set, which we can unwrap by “choosing” its only element. Here we make use of the fact that, although the concept  of unpacking a set does not, in general, exist in the set theory of our  semantic universe, we can mimic it for the case of a singleton set.

\( \, \llbracket \unpack s\rrbracket ^\nu (\rho ) \, = \, choice(\llbracket s\rrbracket ^\nu (\rho )) \, \)

We now consider some further expressions which can be expressed without considering the denotation of predicates.

The maplet rule gives the denotation of $E \, \mapsto  \, F$ for general expressions $E$ and $F$, which may be bunches with multiple elements. The values denoted by $E$ and $F$ in the semantic universe will be sets, and by taking the cross product of these we obtain the set denoting all maplets in bunch $E \, \mapsto  \, F$.

\( \, \llbracket E \, \mapsto  \, F\rrbracket ^\nu (\rho ) \, = \, \llbracket E\rrbracket ^\nu (\rho ) \, × \, \llbracket F\rrbracket ^\nu (\rho ) \, \)

For the denotation of an {\em element} which is a power set, we can allow ourselves to reflect that any elementary values we can represent in our target language of bunch theory will also exist in the set theory language of our semantic universe. Indeed, the term $choice(\llbracket s\rrbracket ^\nu (\rho ))$ will represent, in the semantic universe, exactly the value that we intend $s$ to represent in our target language under environment $\rho $. We call this the “facsimile” of  $s$. We obtain our denotation by taking the power-set of this term 
and enclosing it in set brackets.

\( \, \llbracket \pow (s)\rrbracket ^\nu (\rho ) \, = \, \{ \, \pow (choice(\llbracket s\rrbracket ^\nu (\rho ))) \, \} \, \)

Consider a simple application of the above rule in which $s$ is the constant term $\{1,2\}$. Then using our denotation for constants, we have $\llbracket s\rrbracket ^\nu (\rho ) \, = \, \llbracket \{1,2\}\rrbracket  \, = \, \{\{1,2\}\}$.  Hence $choice(\llbracket s\rrbracket ^\nu (\rho )) \, = \, \{1,2\}$ and:\\
$\{ \, \pow (choice(\llbracket s\rrbracket ^\nu (\rho )))) \, \} \, = \, \{\pow (\{1,2\})\} \, = \, \{ \, \{ \, \{\} \, , \, \{1\} \, , \, \{2 \, , \, \{1,2\} \, \} \, \}$.

For the cross product denotation, we form the terms $choice(\llbracket s\rrbracket ^\nu (\rho ))$ and $choice(\llbracket t\rrbracket ^\nu (\rho ))$ in the semantic universe, and, since these will have the same values as we intend for $s$ and $t$ under environment $\rho $ in the target language, we form the denotation by enclosing the cross product of the former terms in set brackets. 

\( \, \llbracket s \, × \, t\rrbracket ^\nu (\rho ) \, = \, \{ \, choice(\llbracket s\rrbracket ^\nu (\rho )) \, × \, choice(\llbracket t\rrbracket ^\nu (\rho )) \, \} \, \)

For the denotation of $choice(s)$, we identify the special cases in which $s$ has denotation $\{ \, \}$ or $\{\{\}\}$ which correspond to $s$ being $null$ or the empty set respectively. In these cases $choice(s)$ will be $null$ and its denotation will be the empty set. Otherwise  we form the term $choice(\llbracket s\rrbracket ^\nu (\rho ))$ in the semantic universe, since this will have the value we intend for $s$ under environment $\rho $ in our target language. We take the choice of this term and our denotation is the singleton set containing this element. 

$\begin{array}{l} \llbracket choice(s)\rrbracket ^\nu (\rho ) \, = \\
 \,  \, \IF \,  \\
 \,  \,  \,  \,  \,  \,  \, \llbracket s\rrbracket ^\nu (\rho ) \, = \, \{ \, \} \,  \, ∨ \,  \, \llbracket s\rrbracket ^\nu (\rho ) \, = \, \{ \, \{\} \, \} \\
 \,  \, \THEN \,  \\
 \,  \,  \,  \,  \,  \, \{ \, \} \\
 \,  \, \ELSE \\
 \,  \,  \,  \,  \,  \, \{ \, choice(choice(\llbracket s\rrbracket ^\nu (\rho )) \, \} \\
 \,  \, \END \\
\end{array}$

The diligent reader may wish to work through this formula for a simple example.
We will do so for $\llbracket s\rrbracket ^\nu (\rho ) \, = \, \{\{1,2\}\}$, (as would be the case, for example, if $s$ was the constant term $\{1,2\}$) and where $choice\{1,2\} \, = \, 2$. In this case the denotation is given by the $\ELSE$ clause, and we have: 

$\{ \, choice(choice(\llbracket s\rrbracket ^\nu (\rho )) \, \} \, = \, \{ \, choice(choice(\{\{1,2\}\}) \, \} \, = \, \{choice(\{1,2\})\} \, = \, \{2\}$.

We next turn to “truth denotations” of fundamental predicates. Recall that given a set of environments $\mathcal{E} $ and a predicate $P \, $, $ \, \llbracket P\rrbracket ^\tau (\mathcal{E} )$ will be that subset of $\mathcal{E} $ containing environments in  which $P$ is true.

\( \, \llbracket true\rrbracket ^\tau (\mathcal{E} ) \, = \, \mathcal{E}  \,  \,  \,  \,  \,  \, \)
\( \, \llbracket false\rrbracket ^\tau (\mathcal{E} ) \, = \, \{ \, \} \, \)

To express the semantics of equality we form the set comprehension of those environments from $\mathcal{E} $ for which the denotation of $E$ is equal to that of $F$.

\( \, \llbracket E \, = \, F\rrbracket ^\tau (\mathcal{E} ) \, = \, \{ \, \rho  \, | \, \rho  \, ∈ \, \mathcal{E}  \, ∧ \, \llbracket E\rrbracket ^\nu (\rho ) \, = \, \llbracket F\rrbracket ^\nu (\rho ) \, \} \, \)

For set membership $E \, ∈ \, s$, $E$ should consist only of elements which belong to $s$. The denotation of $E$ in some environment $\rho $ will be the set containing these elements, and this must be a subset of the semantic facsimile of $s$.
 
\( \, \llbracket E \, ∈ \, s\rrbracket ^\tau (\mathcal{E} ) \, = \, \{ \, \rho  \, | \, \rho  \, ∈ \, \mathcal{E}  \, ∧ \, \llbracket E\rrbracket ^\nu (\rho ) \, ⊆ \, choice(\llbracket s\rrbracket ^\nu (\rho )) \, \} \, \)

We need a predicate to express that a value is an element. This will be the case when its denotation is a singleton set.

\( \, \llbracket δ(E)\rrbracket ^\tau (\mathcal{E} ) \, = \, \{ \, \rho  \, | \, \rho  \, ∈ \, \mathcal{E}  \, ∧ \, card(\llbracket E\rrbracket ^\nu (\rho )) \, = \, 1 \, \} \, \)

Now composite predicates.

\( \, \llbracket ¬P\rrbracket ^\tau (\mathcal{E} ) \, = \, \{ \, \rho  \, | \, \rho  \, ∈ \, \mathcal{E}  \, ∧ \, \llbracket P\rrbracket ^\tau \{\rho \} \, = \, \{ \, \} \, \} \, \)

\( \, \llbracket P \, ∧ \, Q\rrbracket ^\tau (\mathcal{E} ) \, = \, \llbracket P\rrbracket ^\tau (\mathcal{E} ) \, ∩ \, \llbracket Q\rrbracket ^\tau (\mathcal{E} ) \, \)

\( \, \llbracket P \, ∨ \, Q\rrbracket ^\tau (\mathcal{E} ) \, = \, \llbracket P\rrbracket ^\tau (\mathcal{E} ) \, ∪ \, \llbracket Q\rrbracket ^\tau (\mathcal{E} ) \, \)

For the guarded bunch $g\bguard  \, E$ we need a way to say whether $g$ is true in a particular environment $\rho $. For this we apply the truth semantics of $g$ to the singleton set $\{\rho \}$. Where $g$ is true we have $\llbracket g\rrbracket ^\tau \{\rho \} \, = \, \{\rho \}$.

\( \, \llbracket g\bguard  \, E\rrbracket ^\nu (\rho ) \, = \, \IF \, \llbracket g\rrbracket ^\tau \{\rho \} \, = \, \{\rho \} \, \THEN \, \llbracket E\rrbracket ^\nu (\rho ) \, \ELSE \,  \, \{ \, \} \,  \, \END \, \)

Free variables in our source language may be declared in the forms $\{x \, | \, x \, ∈ \, s \, ∧ \, P\} \, , \, ∀ \, x \, \bullet  \, x \, ∈ \, s \, ⇒ \, P \, , \, ∃ \, x \, \bullet  \, x \, ∈ \, s \, ∧ \, P$. To express the denotation of an occurrence of a free variable $x$ ranging over $s$ we form the facsimile of $s$ in the semantic universe and declare a free variable $x'$ which ranges over the facsimile, and we use binding override $\rho  \, ⊕ \, \lbb  \, x \, \leadsto  \, \{x'\} \, \rbb$ to describe the variable capture in the new environment corresponding to the scope of $x$. The normal semantics for a variable then applies, with: $\llbracket x\rrbracket ^\nu (\rho  \, ⊕ \, \lbb  \, x \, \leadsto  \, \{x'\} \, \rbb) \, = \, \{x'\}$. 

We next consider set comprehension, where we will again need to employ binding override. We form, in the semantic universe, the facsimile of the set we are intending to represent in the source language, and we enclose this 
within another layer of set brackets. To produce the facsimile of 

$\{ \, x \, | \, x \, ∈ \, s \, ∧ \, P \, \}$ 

we use a set comprehension in which bound variable $x'$ ranges over the facsimile of $s$, and we retain those $x'$ for which our intention is that predicate $P$ is true in environment $\rho $ overridden with a binding of $x$ to $\{x'\}$. The trick for expressing this is to apply the truth semantics of $P$ to a singleton set of environments containing just $\rho $ with the binding for $x$ overridden by $\lbb x \, \leadsto  \, \{x'\}\rbb$. If this yields an empty set, $P$ is false in $\rho $, otherwise it is true.

$\begin{array}{l} \llbracket \{ \, x \, | \, x \, ∈ \, s \, ∧ \, P\rrbracket ^\nu (\rho ) \, = \,  \\
 \,  \,  \,  \,  \,  \,  \,  \,  \,  \,  \, \{ \,  \, \{ \, x' \, | \, x' \, ∈ \, choice(\llbracket s\rrbracket ^\nu (\rho )) \, ∧ \, \llbracket P\rrbracket ^\tau (\{\rho  \, ⊕ \, \lbb  \, x \, \leadsto  \, \{x'\} \, \rbb\})≠\{ \, \} \, \} \,  \, \} \\
\end{array}$

The truth denotations for universally and existentially quantified predicates use similar techniques to those employed for set comprehension.

$\begin{array}{l} \llbracket ∀x \, \bullet  \, x \, ∈ \, s \, ⇒ \, P\rrbracket ^\tau (\mathcal{E} ) \, = \, \{ \, \rho  \, | \, \rho  \, ∈ \, \mathcal{E}  \, ∧ \, (∀ \, x'\bullet  \, x' \, ∈ \, choice(\llbracket s\rrbracket ^\nu (\rho )) \,  \\
 \,  \,  \,  \,  \,  \,  \,  \,  \,  \,  \,  \,  \,  \,  \,  \, ⇒ \,  \, \llbracket P\rrbracket ^\tau (\{\rho  \, ⊕ \, \lbb  \, x \, \leadsto  \, \{x'\} \, \rbb\})≠\{ \, \}) \, \} \\
\end{array}$

$\begin{array}{l} \llbracket ∃x \, \bullet  \, x \, ∈ \, s \, ∧ \, P\rrbracket ^\tau (\mathcal{E} ) \, = \, \{ \, \rho  \, | \, \rho  \, ∈ \, \mathcal{E}  \, ∧ \, (∃ \, x'\bullet  \, x' \, ∈ \, choice(\llbracket s\rrbracket ^\nu (\rho )) \\
 \,  \,  \,  \,  \,  \,  \,  \,  \,  \,  \,  \,  \,  \,  \,  \, ∧ \, \llbracket P\rrbracket ^\tau (\{\rho  \, ⊕ \, \lbb  \, x \, \leadsto  \, \{x'\} \, \rbb\})≠\{ \, \}) \, \} \,  \\
\end{array}$

Where the predicate $P$ is $true$, the denotation for the existential predicate becomes:

\( \, \llbracket ∃ \, x \, \bullet  \, x \, ∈ \, s\rrbracket ^\tau (\mathcal{E} ) \, = \, \{ \, \rho  \, | \, \rho  \, ∈ \, \mathcal{E}  \, ∧ \, (∃ \, x' \, \bullet  \, x' \, ∈ \, choice(\llbracket s\rrbracket ^\nu (\rho ))) \, \} \, \)

We now consider the denotations of expressions and predicates subject to substitution of certain terms. For these, we capture the effect of the substitution through binding override.

\( \, \llbracket E[F/x]\rrbracket ^\nu (\rho ) \, = \, \llbracket E\rrbracket ^\nu (\rho  \, ⊕ \, \lbb  \, x \, \leadsto  \, \llbracket F\rrbracket ^\nu (\rho ) \, \rbb) \, \)

\( \, \llbracket P[F/x]\rrbracket ^\tau (\mathcal{E} ) \, = \, \{ \, \rho  \, | \, \rho  \, ∈ \, \mathcal{E}  \, ∧ \, \llbracket P\rrbracket ^\tau (\{\rho  \, ⊕ \, \lbb  \, x \, \leadsto  \, \llbracket F\rrbracket ^\nu (\rho ) \, \rbb\}) \, ≠ \, \{\} \, \} \, \)

Our final two rules give the denotation of the set $BIG$ and the semantic characterisation of an infinite set. We use the naturals to construct the denotation of $BIG$. The idea here is just to provide an infinite set for the construction of mathematical objects. 

\( \, \llbracket BIG\rrbracket ^\nu (\rho ) \, = \, \{ \, \nat  \, \} \, \)

The semantic characterisation of an infinite set requires its facsimile in the semantic model to be infinite.

\( \, \llbracket infinite(s)\rrbracket ^\tau (\mathcal{E} ) \, = \, \{ \, \rho  \, | \, \rho  \, ∈ \, \mathcal{E}  \, ∧ \, infinite(choice(\llbracket s\rrbracket ^\nu (\rho ))) \, \} \, \)

\section{Proving the existence of a model}

So far we have given axioms for bunch theory, and given semantic rules for  translating any expression or predicate in the basic language of bunch theory into the language of our semantic universe. 

At this point, the key question is whether the semantic translations of the axioms of bunch theory yield predicates which can be proved to be true in the semantic domain. If we can establish this we will have proved that a model exists for our version of bunch theory.

All expressions in our source language which are intended to denote elements will have denotations which are singleton sets. The formal proof of this is by structural induction over the syntactic forms of the source language, appealing to the semantic rules, and with the base cases being for constants and variables.

Where an axiom has the form $P \, ⇔ \, Q$ we validate it within the model by  showing \( \, \llbracket P\rrbracket ^\tau (\mathcal{E} ) \, = \, \llbracket Q\rrbracket ^\tau (\mathcal{E} ) \, \) for arbitrary $\mathcal{E} $.

Where an axiom has the form $P \, ⇒ \, Q$ we validate it by showing \( \, \llbracket P\rrbracket ^\tau (\mathcal{E} ) \, ⊆ \, \llbracket Q\rrbracket ^\tau (\mathcal{E} \} \, \) for arbitrary $\mathcal{E} $.

Where an axiom has the form $E=F$ we validate it by showing \( \, \llbracket E\rrbracket ^\nu (\rho ) \, = \, \llbracket F\rrbracket ^\nu (\rho ) \, \) for arbitrary $\rho $.

Where an axiom has the general form $P$ we validate it by showing \(\llbracket P\rrbracket ^\tau (\mathcal{E} ) \, = \, \mathcal{E} \) for arbitrary $\mathcal{E} $.

To validate an inference rule we show that any environment that results in the premises of  the rule being true will also result in the  conclusion of the rule being true.

For our validation we use the set theory formulated in the B-Book\cite{JRA96}, augmented with environments.  During the validations we call on various properties of this set theory, including its axioms.

We give illustrative examples of axiom validations. An more extensive validation is given in \cite{arXivBunches_v2}.

\subsection{Example validations of set theory axioms}

It is no surprise to find that our set theory axioms can be validated; we already have faith in set theory! However, the validation provides assurance that our model is correct. We give example validations for the ordered pairs and set comprehension axioms.

\subsubsection{Validation of the ordered pair axiom}

To validate \( \, E\mapsto F \, ∈ \, (s \, × \, t) \,  \, ⇔ \,  \, E \, ∈ \, s \,  \, ∧ \,  \, F \, ∈ \, t\)

We show: 

\( \, \llbracket  \, E\mapsto F \, ∈ \, (s \, × \, t)\rrbracket ^\tau (\mathcal{E} ) \,  \, = \,  \, \llbracket E \, ∈ \, s \,  \, ∧ \,  \, F \, ∈ \, t\rrbracket ^\tau (\mathcal{E} ) \, \)

Taking the LHS

$\begin{array}{l} \llbracket E\mapsto F \, ∈ \, (s \, × \, t)\rrbracket ^\tau (\mathcal{E} ) \, = \, \q{by semantics of set membership} \end{array}$

$\begin{array}{l}
\{ \, \rho  \, | \, \rho  \, ∈ \, \mathcal{E}  \,  \, ∧ \,  \, \llbracket E\mapsto F\rrbracket ^\nu (\rho ) \, ⊆ \, choice(\llbracket s \, × \, t\rrbracket ^\nu (\rho )) \, \} \end{array}$

$\begin{array}{l}
= \, \q{by semantics of ordered pair and cross product} \end{array}$

$\begin{array}{l}
\{ \, \rho  \, | \, \rho  \, ∈ \, \mathcal{E}  \,  \, ∧ \,  \,  \\
 \,  \,  \, \llbracket E\rrbracket ^\nu (\rho ) \, × \, \llbracket F\rrbracket ^\nu (\rho ) \, ⊆ \, choice(\{choice(\llbracket s\rrbracket ^\nu (\rho ) \, × \, choice(\llbracket t\rrbracket ^\nu (\rho )\}) \,  \\
\} \end{array}$

$\begin{array}{l}
= \, \text{“applying choice and by property that for non-empty $a$ and $b$,} \,  \end{array}$

$\begin{array}{l}
a×b \, ⊆ \,  \, c×d \,  \, ⇔ \,  \, a⊆c \, ∧ \,  \, b⊆d\text{”} \end{array}$

$\begin{array}{l}
\{ \, \rho  \, | \, \rho  \, ∈ \, \mathcal{E}  \,  \, ∧ \,  \,  \\
 \,  \,  \, \llbracket E\rrbracket ^\nu (\rho ) \, ⊆ \, choice(\llbracket s\rrbracket ^\nu (\rho )) \,  \, ∧ \,  \, \llbracket F\rrbracket ^\nu (\rho ) \, ⊆ \, choice(\llbracket t\rrbracket ^\nu (\rho )) \,  \\
\} \end{array}$

$\begin{array}{l}
= \, \q{by property of set intersection} \end{array}$

$\begin{array}{l}
\{ \, \rho  \, | \, \rho  \, ∈ \, \mathcal{E}  \,  \, ∧ \,  \, \llbracket E\rrbracket ^\nu (\rho ) \, ⊆ \, choice(\llbracket s\rrbracket ^\nu (\rho )) \, \} \,  \\
 \,  \,  \, ∩ \,  \\
\{ \, \rho  \, | \, \rho  \, ∈ \, \mathcal{E}  \,  \, ∧ \,  \, \llbracket F\rrbracket ^\nu (\rho ) \, ⊆ \, choice(\llbracket t\rrbracket ^\nu (\rho )) \, \} \end{array}$

$\begin{array}{l}
= \, \q{by semantics of set membership} \end{array}$

$\begin{array}{l}
\llbracket E \, ∈ \, s\rrbracket ^\tau (\mathcal{E} ) \, ∩ \, \llbracket F \, ∈ \, t\rrbracket ^\tau (\mathcal{E} ) \,  \, = \,  \, \q{by semantics of $∧$} \end{array}$

$\begin{array}{l}
\llbracket E \, ∈ \, s \,  \, ∧ \,  \, F \, ∈ \, t\rrbracket ^\tau (\mathcal{E} ) \,  \,  \,  \,  \, _\square  \,  \,  \,  \\
\end{array}$

Here we have validated a more general result than the original 
ordered pair axiom, since $E$ and $F$ in the above may be general 
expressions rather than expressions that must represent elements.

\subsubsection{Validation of set comprehension axiom}

\(E \, ∈ \, \{x|x \, ∈ \, s \, ∧ \, P\} \,  \, ⇔ \,  \, E \, ∈ \, s \,  \, ∧ \,  \, P[E/x] \,  \,  \,  \,  \,  \,  \, \text{where} \,  \, x∖s \,  \, \)

We show:

\( \, \llbracket E \, ∈ \, \{x|x \, ∈ \, s \, ∧ \, P\}\rrbracket ^\tau (\mathcal{E} ) \, = \, \llbracket E \, ∈ \, s \, ∧ \, P[E/x]\rrbracket ^\tau (\mathcal{E} ) \, \)

Taking LHS

$\begin{array}{l} \llbracket E \, ∈ \, \{ \, x \, | \, x \, ∈ \, s \, ∧ \, P\}\rrbracket ^\tau (\mathcal{E} ) \, = \, \q{by semantics of set membership} \end{array}$

$\begin{array}{l}
\{ \, \rho  \, | \, \rho  \, ∈ \, \mathcal{E}  \, ∧ \, \llbracket E\rrbracket ^\nu (\rho ) \, ⊆ \, choice(\llbracket \{ \, x \, | \, x \, ∈ \, s \, ∧ \, P\}\rrbracket ^\nu (\rho )) \, \} \end{array}$

$\begin{array}{l}
= \, \q{by semantics of set comprehension} \end{array}$

$\begin{array}{l}
\{ \, \rho  \, | \, \rho  \, ∈ \, \mathcal{E}  \, ∧ \, \llbracket E\rrbracket ^\nu (\rho ) \, ⊆ \,  \\
 \,  \,  \, choice( \\
 \,  \,  \,  \,  \,  \, \{ \, \{ \, x' \, | \, x' \, ∈ \, choice(\llbracket s\rrbracket ^\nu (\rho ) \, ∧ \, \llbracket P\rrbracket ^\tau (\{\rho  \, ⊕ \, \lbb  \, x \, \leadsto  \, \{x'\} \, \rbb\}) \, ≠ \, \{\} \, \} \, \} \\
 \,  \,  \, ) \,  \\
\} \,  \end{array}$

$\begin{array}{l}
= \, \q{taking choice of singleton set} \,  \,  \end{array}$

$\begin{array}{l}
\{ \, \rho  \, | \, \rho  \, ∈ \, \mathcal{E}  \, ∧ \, \llbracket E\rrbracket ^\nu (\rho ) \, ⊆ \,  \\
 \,  \,  \, \{ \, x' \, | \, x' \, ∈ \, choice(\llbracket s\rrbracket ^\nu (\rho )) \, ∧ \, \llbracket P\rrbracket ^\tau (\{\rho  \, ⊕ \, \lbb  \, x \, \leadsto  \, \{x'\} \, \rbb\}) \, ≠ \, \{\} \, \} \,  \,  \,  \\
\} \end{array}$

$\begin{array}{l}
= \, \text{“by singleton set property} \, \{a\} \, ⊆ \, B \,  \, ⇔ \,  \, a \, ∈ \, B \, \text{”} \end{array}$

$\begin{array}{l}
\{ \, \rho  \, | \, \rho  \, ∈ \, \mathcal{E}  \, ∧ \, choice(\llbracket E\rrbracket ^\nu (\rho )) \, ∈ \\
 \,  \,  \, \{ \, x' \, | \, x' \, ∈ \, choice(\llbracket s\rrbracket ^\nu (\rho )) \, ∧ \, \llbracket P\rrbracket ^\tau (\{\rho  \, ⊕ \, \lbb  \, x \, \leadsto  \, \{x'\} \, \rbb\}) \, ≠ \, \{\} \, \} \\
\} \end{array}$

$\begin{array}{l}
\text{Now recalling set theory axiom 3:} \end{array}$

$\begin{array}{l}
E \, ∈ \, \{ \, x \, | \, x \, ∈ \, s \, ∧ \, P \, \} \,  \, ⇔ \,  \, E \, ∈ \, s \,  \, ∧ \,  \, P[E/x] \end{array}$

$\begin{array}{l}
\text{applying this to} \end{array}$

$\begin{array}{l}
choice(\llbracket E\rrbracket ^\nu (\rho )) \, ∈ \,  \\
 \,  \,  \, \{ \, x' \, | \, x' \, ∈ \, choice(\llbracket s\rrbracket ^\nu (\rho )) \, ∧ \, \llbracket P\rrbracket ^\tau (\{\rho  \, ⊕ \, \lbb  \, x \, \leadsto  \, \{x'\} \, \rbb\}) \, ≠ \, \{\} \, \} \end{array}$

$\begin{array}{l}
\text{with:} \end{array}$

$\begin{array}{l}
choice(\llbracket E\rrbracket ^\nu (\rho )) \,  \, \text{for} \,  \,  \, E \end{array}$

$\begin{array}{l}
x' \,  \, \text{for} \,  \, x \end{array}$

$\begin{array}{l}
choice(\llbracket s\rrbracket ^\nu (\rho )) \,  \, \text{for} \,  \, s \end{array}$

$\begin{array}{l}
\text{and } \, \llbracket P\rrbracket ^\tau (\{\rho  \, ⊕ \, \lbb  \, x \, \leadsto  \, \{x'\} \, \rbb\}) \, ≠ \, \{\} \,  \, \text{for} \,  \, P \end{array}$

$\begin{array}{l}
\text{the predicate } \, E \, ∈ \, s \,  \, ∧ \,  \, P[E/x] \, \text{ in the axiom becomes} \end{array}$

$\begin{array}{l}
choice(\llbracket E\rrbracket ^\nu (\rho )) \, ∈ \, choice(\llbracket s\rrbracket ^\nu (\rho )) \,  \, ∧ \\
(\llbracket P\rrbracket ^\tau (\{\rho  \, ⊕ \, \lbb  \, x \, \leadsto  \, \{x'\} \, \rbb\}) \, ≠ \, \{\}) \, [choice(\llbracket E\rrbracket ^\nu (\rho )) \, / \, x'] \end{array}$

$\begin{array}{l}
\text{which after performing the indicated substitution becomes} \end{array}$

$\begin{array}{l}
choice(\llbracket E\rrbracket ^\nu (\rho )) \, ∈ \, choice(\llbracket s\rrbracket ^\nu (\rho )) \,  \, ∧ \\
\llbracket P\rrbracket ^\tau (\{\rho  \, ⊕ \, \lbb  \, x \, \leadsto  \, \{choice(\llbracket E\rrbracket ^\nu (\rho ))\} \, \rbb\}) \, ≠ \, \{\} \end{array}$

$\begin{array}{l}
\text{And thus, LHS} \,  \, = \end{array}$

$\begin{array}{l}
\{ \, \rho  \, | \, \rho  \, ∈ \, \mathcal{E}  \, ∧ \, choice(\llbracket E\rrbracket ^\nu (\rho )) \, ∈ \, choice(\llbracket s\rrbracket ^\nu (\rho )) \,  \,  \\
∧ \, \llbracket P\rrbracket ^\tau (\{\rho  \, ⊕ \, \lbb  \, x \, \leadsto  \, \{choice(\llbracket E\rrbracket ^\nu (\rho ))\} \, \rbb\}) \, ≠ \, \{\} \, \} \\
\end{array}$

We now take RHS and show it equal to the same expression:

$\begin{array}{l} \llbracket E \, ∈ \, s \,  \, ∧ \,  \, P[E/x]\rrbracket ^\tau (\mathcal{E} ) \, = \, \q{by semantics of $∧$} \end{array}$

$\begin{array}{l}
\llbracket E \, ∈ \, s\rrbracket ^\tau (\mathcal{E} ) \,  \, ∩ \,  \, \llbracket P[E/x]\rrbracket ^\tau (\mathcal{E} ) \end{array}$

$\begin{array}{l}
= \, \q{by semantic rules for set membership and substitution} \end{array}$

$\begin{array}{l}
\{ \, \rho  \, | \, \rho  \, ∈ \, \mathcal{E}  \,  \, ∧ \,  \, \llbracket E\rrbracket ^\nu (\rho ) \, ⊆ \, choice(\llbracket s\rrbracket ^\nu (\rho )) \, \} \,  \,  \, \cap \\
\{ \, \rho  \, | \, \rho  \, ∈ \, \mathcal{E}  \,  \, ∧ \,  \, \llbracket P\rrbracket ^\tau (\{\rho  \, ⊕ \, \lbb  \, x \, \leadsto  \, \llbracket E\rrbracket ^\nu (\rho ) \, \rbb\}) \, ≠ \, \{\} \, \} \,  \,  \end{array}$

$\begin{array}{l}
= \, \q{by property of set intersection} \end{array}$

$\begin{array}{l}
\text{LHS} \,  \, _\square  \\
\end{array}$

\subsection{Validation of the axioms specific to bunch theory}

The validation of the axioms specific to bunch theory establishes that they are true (with respect to the model).

\subsubsection{Bunch theory axiom, packaging 1}

$\begin{array}{l} \{\unpack A\} \, = \, A \,  \,  \, \text{where} \,  \, A \,  \, \text{is an element and a set.} \\
\end{array}$

To validate we show:

\( \, \llbracket \{\unpack A\}\rrbracket ^\nu (\rho ) \, = \, \llbracket A\rrbracket ^\nu (\rho ) \, \)

Taking LHS

$\begin{array}{l} \llbracket \{\unpack A\}\rrbracket ^\nu (\rho ) \, = \, \q{by semantics of packaging} \end{array}$

$\begin{array}{l}
\{\llbracket \unpack A\rrbracket ^\nu (\rho )\} \, = \, \q{by semantics of unpackaging} \end{array}$

$\begin{array}{l}
\{choice(\llbracket A\rrbracket ^\nu (\rho )\} \, = \, \q{since the choice is from a singleton set} \end{array}$

$\begin{array}{l}
\llbracket A\rrbracket ^\nu (\rho ) \,  \,  \,  \,  \,  \, _\square  \\
\end{array}$

\subsubsection{Bunch theory axiom, packaging 2}
$\begin{array}{l} \unpack \{E\} \, = \, E \\
\end{array}$

To validate we show:

\( \, \llbracket \unpack \{E\}\rrbracket ^\nu (\rho ) \, = \, \llbracket E\rrbracket ^\nu (\rho ) \, \)

Taking LHS

$\begin{array}{l} \llbracket \unpack \{E\}\rrbracket ^\nu (\rho ) \, = \, \q{by semantics of unpackaging} \end{array}$

$\begin{array}{l}
choice(\llbracket \{E\}\rrbracket ^\nu (\rho )) \, = \, \q{by semantics of packaging} \end{array}$

$\begin{array}{l}
choice(\{\llbracket E\rrbracket ^\nu (\rho )\}) \, = \, \q{taking choice} \end{array}$

$\begin{array}{l}
\llbracket E\rrbracket ^\nu (\rho ) \,  \,  \,  \,  \,  \,  \,  \, _\square  \\
\end{array}$

\subsubsection{Bunch theory element axiom 1}

\( \, δ(\unpack A) \,  \, ⇔ \,  \, A \, = \, \{choice(A)\} \,  \,  \, \)  ($A$ an element and a set.)

To validate we show:

\( \, \llbracket δ(\unpack A)\rrbracket ^\tau (\mathcal{E} ) \,  \, = \,  \, \llbracket A \, = \, \{choice(A)\}\rrbracket ^\tau (\mathcal{E} ) \, \)

$\eq \,  \, $ “by semantics of $δ(E)$ on LHS and $=$ on RHS”

 \( \, \{ \, \rho  \, | \, \rho  \, ∈ \, \mathcal{E}  \,  \, ∧ \,  \, card(\llbracket \unpack A\rrbracket ^\nu (\rho ) \, = \, 1 \, \} \, = \,  \\
\{ \, \rho  \, | \, \rho  \, ∈ \, \mathcal{E}  \,  \, ∧ \, \llbracket A\rrbracket ^\nu (\rho ) \, = \, \llbracket \{choice(A)\}\rrbracket ^\nu (\rho ) \, \} \, \)

$\eq \, $ “by property of set comprehension”

\( \, card(\llbracket \unpack A\rrbracket ^\nu (\rho ) \, = \, 1 \,  \, ⇔ \,  \,  \, \llbracket A\rrbracket ^\nu (\rho ) \, = \, \llbracket \{choice(A)\}\rrbracket ^\nu (\rho ) \, \)

$\eq \, $ “by semantics of unpacking on LHS, with $A$ an element, and semantics of packing on RHS” 

\( \, card(choice(\llbracket A\rrbracket ^\nu (\rho )) \, = \, 1 \, ⇔ \,  \, \llbracket A\rrbracket ^\nu (\rho ) \, = \, \{\llbracket choice(A)\rrbracket ^\nu (\rho )\} \, \)

$\eq \, $ “applying semantics of choice to RHS”

\( \, card(choice(\llbracket A\rrbracket ^\nu (\rho )) \, = \, 1 \, ⇔ \,  \, \llbracket A\rrbracket ^\nu (\rho ) \, = \,  \\
\{\IF \, ∃ \, x' \, \bullet  \, x' \, ∈ \, choice(\llbracket A\rrbracket ^\nu (\rho )) \, \THEN \, \{choice(choice(\llbracket A\rrbracket ^\nu (\rho ))\} \, \ELSE \, \{ \, \} \, \END\} \\
\)

Noting the case $¬∃ \, x' \, \bullet  \, x' \, ∈ \, choice(\llbracket A\rrbracket ^\nu (\rho )$ does not apply since $A$ is an element and therefore in non-null and has a non-empty denotation, our proof goal becomes:

\( \, card(choice(\llbracket A\rrbracket ^\nu (\rho )) \, = \, 1 \, ⇔ \,  \, \llbracket A\rrbracket ^\nu (\rho ) \, = \, \{\{choice(choice(\llbracket A\rrbracket ^\nu (\rho ))\}\} \, \)

Since $A$ is an element its denotation will be a singleton set, and we can rewrite $\llbracket A\rrbracket ^\nu (\rho )$ as $\{choice\llbracket A\rrbracket ^\nu (\rho )\}$, giving:

\( \, card(choice(\llbracket A\rrbracket ^\nu (\rho )) \, = \, 1 \, ⇔ \,  \, \{choice\llbracket A\rrbracket ^\nu (\rho )\} \, = \, \{\{choice(choice(\llbracket A\rrbracket ^\nu (\rho ))\}\} \, \)

Writing $S$ for $choice(\llbracket A\rrbracket ^\nu (\rho )$ we have:

\( \, card(S) \, = \, 1 \, ⇔ \,  \, \{S\} \, = \, \{\{choice(S)\}\} \, \)

\( \, \eq \,  \, card(S) \, = \, 1 \, ⇔ \,  \, S \, = \, \{choice(S)\} \, \)

\( \, \eq \, true \,  \,  \, \)

\subsubsection{Bunch theory element axiom 2}

\( \, δ(\{E\}) \, \)

To validate we show:

\( \, \llbracket δ(\{E\})\rrbracket ^\tau (\mathcal{E} ) \, = \, \mathcal{E}  \, \)

$\begin{array}{l} \llbracket δ(\{E\})\rrbracket ^\tau (\mathcal{E} ) \, \eq \, \q{by semantics of $element$ } \end{array}$

$\begin{array}{l}
\{ \, \rho  \, | \, \rho  \, ∈ \, \mathcal{E}  \, ∧ \, card(\llbracket \{E\}\rrbracket )^\nu (\rho )) \, = \, 1 \, \} \,  \, \eq \, \q{by semantics of packaging} \end{array}$

$\begin{array}{l}
\{ \, \rho  \, | \, \rho  \, ∈ \, \mathcal{E}  \, ∧ \, card(\{\llbracket E\rrbracket ^\nu (\rho )\}) \, = \, 1 \, \} \,  \, \eq \end{array}$

$\begin{array}{l}
\mathcal{E}  \\
\end{array}$

\subsubsection{First guarded bunch axiom}

\( \, g \,  \, ⇒ \,  \, g\bguard  \, E \, = \, E \, \)

To validate we show:

\( \, \llbracket g\rrbracket ^\tau (\mathcal{E} ) \,  \, ⊆ \,  \, \llbracket g\bguard  \, E \, = \, E\rrbracket ^\tau (\mathcal{E} ) \, \)

$\eq$  by set expansion of LHS of $⊆$ and applying equality semantics to RHS.

\( \, \{ \, \rho  \, | \, \rho  \, ∈ \, \mathcal{E}  \, ∧ \, \llbracket g\rrbracket ^\tau \{\rho \} \, = \, \{\rho \} \, \} \, ⊆ \,  \\
 \,  \,  \,  \,  \, \{ \, \rho  \, | \, \rho  \, ∈ \, \mathcal{E}  \,  \, ∧ \, \llbracket g\bguard  \, E\rrbracket ^\nu (\rho ) \, = \,  \, \llbracket E\rrbracket ^\nu (\rho ) \, \} \, \)

$\eq$ by a property of set comprehension
 
\( \, \rho  \, ∈ \, \mathcal{E}  \, ∧ \, \llbracket g\rrbracket ^\tau \{\rho \} \, = \, \{\rho \} \,  \, ⇒ \,  \, \rho  \, ∈ \, \mathcal{E}  \,  \, ∧ \, \llbracket g\bguard  \, E\rrbracket ^\nu (\rho ) \, = \,  \, \llbracket E\rrbracket ^\nu (\rho ) \, \)

We must prove the RHS of the implication under assumptions

$\rho  \, ∈ \, \mathcal{E} $ and $ \, \llbracket g\rrbracket ^\tau \{\rho \} \, = \, \{\rho \}$

By semantics of the guarded bunch, the RHS becomes:

\( \, \rho  \, ∈ \, \mathcal{E}  \,  \, ∧ \,  \,  \\
\IF \, \llbracket g\rrbracket ^\tau \{\rho \} \, = \, \{\rho \} \, \THEN \, \llbracket E\rrbracket ^\nu (\rho ) \, = \, \llbracket E\rrbracket ^\nu (\rho ) \, \ELSE \, \{ \, \} \, \END \, = \, \llbracket E\rrbracket ^\nu (\rho ) \, \)

now $\rho  \, ∈ \, \mathcal{E} $ is already assumed and using assumption $ \, \llbracket g\rrbracket ^\tau \{\rho \} \, = \, \{\rho \}$ to resolve the conditional statement it remains to prove:

\( \, \llbracket E\rrbracket ^\nu (\rho ) \, = \, \llbracket E\rrbracket ^\nu (\rho ) \, \)

Which is true by the equality law we inherit from predicate logic with equality.

\subsubsection{Second guarded bunch axiom}

\( \, ¬g \,  \, ⇒ \,  \, g\bguard  \, E \, = \, \unpack \{ \, \} \, \)

To validate we show:

\( \, \llbracket ¬g\rrbracket ^\tau (\mathcal{E} ) \,  \, ⊆ \,  \, \llbracket ¬g\bguard  \, E \, = \, \unpack \{ \, \}\rrbracket ^\tau (\mathcal{E} ) \,  \, \)   

$\eq$ “by applying semantics of negation to LHS of $⊆$ and equality semantics to RHS”

\( \, \{ \, \rho  \, | \, \rho  \, ∈ \, \mathcal{E}  \, ∧ \, \llbracket g\rrbracket ^\tau \{\rho \} \, = \, \{ \, \} \, \} \, ⊆ \,  \\
 \,  \,  \,  \,  \, \{ \, \rho  \, | \, \rho  \, ∈ \, \mathcal{E}  \,  \, ∧ \, \llbracket ¬g\bguard  \, E\rrbracket ^\nu (\rho ) \, = \,  \, \llbracket \unpack \{\}\rrbracket ^\nu (\rho ) \, \} \, \)

$\eq$ “by property of set comprehension and semantics of unpacking”
 
\( \, \rho  \, ∈ \, \mathcal{E}  \, ∧ \, \llbracket g\rrbracket ^\tau \{\rho \} \, = \, \{ \, \} \,  \, ⇒ \,  \, \rho  \, ∈ \, \mathcal{E}  \,  \, ∧ \, \llbracket g\bguard  \, E\rrbracket ^\nu (\rho ) \, = \,  \, \{ \, \} \, \)

We must prove the RHS of the implication under assumptions

$\rho  \, ∈ \, \mathcal{E} $ and $ \, \llbracket g\rrbracket ^\tau \{\rho \} \, = \, \{ \, \}$

By semantics of the guarded bunch, the RHS becomes:

\( \, \rho  \, ∈ \, \mathcal{E}  \,  \, ∧ \,  \,  \\
\IF \, \llbracket g\rrbracket ^\tau \{\rho \} \, = \, \{\rho \} \, \THEN \, \llbracket E\rrbracket ^\nu (\rho ) \,  \, \ELSE \, \{ \, \} \, \END \, = \, \{ \, \} \, \)

now $\rho  \, ∈ \, \mathcal{E} $ is already assumed and using assumption $ \, \llbracket g\rrbracket ^\tau \{\rho \} \, = \, \{ \, \}$ to resolve the conditional statement it remains to prove:

\( \, \{ \, \} \, = \, \{ \, \} \, \)  which is true by equality.

\subsection{Establishing the soundness of our logic}

The soundness of a sequent $P \, ⊢ \, Q$ is established by demonstrating the corresponding semantic entailment $P \, \vDash  \, Q$ which holds when $\llbracket P\rrbracket ^\tau (\mathcal{E} ) \, ⊆ \, \llbracket Q\rrbracket ^\tau (\mathcal{E} )$

The soundness of a proof rule:

$\begin{array}{l} \displaystyle\frac{HYP \, ⊢ \, P}{HYP \, ⊢ \, Q} \\
\end{array}$

is established by demonstrating $(HYP \, \vDash  \, P) \, ⇒ \, (HYP \, \vDash  \, Q)$ whichj is equivalent to $\llbracket P\rrbracket ^\tau (\mathcal{E} ) \, ⊆ \, \llbracket Q\rrbracket ^\tau (\mathcal{E} )$

Other proof rules can be massaged into this form bearing in mind that the sequent $P,Q \, ⊢ \, R$ is equivalent to $P \, ⊢ \, Q⇒R$ and that where a proof rule has as its premises the two sequents $P \, ⊢ \, Q$ and $P \, ⊢ \, R$ these can be replaced by the single premise $P \, ⊢ \, Q∧R$.

$\exists$

\subsubsection{The soundness of the exists-intro rule of FOPLN}

Our rule is:

 $\begin{array}{l} \displaystyle\frac{HYP \, ⊢ \, δ(E) \, , \,  \, HYP \, ⊢ \,  \, P[E/X]}{HYP \, ⊢ \, ∃ \, x \, \bullet  \, P} \,  \, \text{       ∃-intro} \\
\end{array}$  

By the arguments given above it will be sufficient to prove:

$\begin{array}{l} \llbracket δ(E)\rrbracket ^\tau (\mathcal{E} ) \, ∩ \, \llbracket P[E/x]\rrbracket ^\tau (\mathcal{E} ) \, ⊆ \, \llbracket ∃x\bullet P\rrbracket ^\tau (\mathcal{E} ) \\
\text{which by the definitions of} \, \llbracket δ(E)\rrbracket ^\tau (\mathcal{E} ), \, \llbracket P[E/x]\rrbracket ^\tau (\mathcal{E} ) \, \text{and} \, \llbracket ∃x\bullet P\rrbracket ^\tau (\mathcal{E} ) \\
\text{is equivalent to} \\
\{\rho  \, | \, \rho ∈\mathcal{E}  \, ∧ \, \card(\llbracket E\rrbracket ^\nu (\rho ))=1\} \, ∪ \, \{\rho  \, | \, \rho  \, ∈ \, \mathcal{E}  \, ∧ \, (∃x' \, \bullet  \, \llbracket P\rrbracket ^\tau (\{\rho ⊕\lbb x\leadsto \{x'\}\rbb\}) \, ≠ \, \{\} \, \} \\
⊆ \, \{ \, \rho  \, | \, \rho  \, ∈ \, \mathcal{E}  \, ∧ \, (∃ \, x' \, \bullet  \, \llbracket P\rrbracket ^\tau (\{\rho ⊕\lbb x\leadsto \{x'\}\rbb\}) \,  \, ≠ \, \{\} \, \} \\
\eq \, \text{extracting predicates} \\
 \, \rho ∈\mathcal{E}  \, ∧ \, card(\llbracket E\rrbracket ^\nu (\rho ))=1 \, ∧ \, \llbracket P\rrbracket ^\tau (\{\rho ⊕\lbb x \, \leadsto \llbracket E\rrbracket ^\nu (\rho )\rbb\}) \, ≠ \, \{\} \\
 \,  \,  \, ⇒ \\
\rho  \, ∈ \, \mathcal{E}  \, ∧ \, (∃x' \, \bullet  \, \llbracket P\rrbracket ^\tau (\{\rho ⊕\lbb x\leadsto \{x'\}\rbb\}) \, ≠ \, \{\}) \\
\text{which  we can abstrat as the form } \, a∧b∧c \, ⇒ \, a∧d \\
\text{and noting from logic that} \, (a∧b∧c \, ⇒ \, a∧d \,  \, ⇔ \,  \, (a \, ⇒ \, (b∧c \, ⇒ \, d)) \,  \\
\text{we can write our proof goal as:} \\
\rho ∈\mathcal{E}  \, ⇒ \,  \\
 \,  \, (card(\llbracket E\rrbracket ^\nu (\rho ) \, = \, 1 \, ∧ \, \llbracket P\rrbracket ^\tau (\{\rho ⊕\lbb x\leadsto \llbracket E\rrbracket ^\nu \rbb\}) \, ≠ \, 1 \,  \\
 \,  \,  \,  \,  \,  \, ⇒ \\
 \,  \, (∃x' \, \bullet  \, \llbracket P\rrbracket ^\tau (\{\rho ⊕\lbb x\leadsto \{x'\}\rbb\}) \, ≠ \, \{\}) \,  \, ) \end{array}$

$\begin{array}{l}
\text{which, since }b \, ⇒ \, (a \, ⇒ \, b) \, \text{is implied by} \end{array}$

$\begin{array}{l}
card(\llbracket E\rrbracket ^\nu (\rho ) \, = \, 1 \, ∧ \, \llbracket P\rrbracket ^\tau (\{\rho ⊕\lbb x\leadsto \llbracket E\rrbracket ^\nu \rbb\}) \, ≠ \, \{\} \\
 \,  \,  \,  \,  \,  \, ⇒ \\
(∃x' \, \bullet  \, \llbracket P\rrbracket ^\tau (\{\rho ⊕\lbb x\leadsto \{x'\}\rbb\}) \, ≠ \, \{\}) \,  \, ) \\
\end{array}$

We now appeal to the one point rule: $(∃x \, \bullet x \, = \, E \, ∧ \, P) \, ⇔ \, P[E/x]$

with $x' \, = \, choice(\llbracket E\rrbracket ^\nu (\rho ))$. Since from LHS of our implication we  have $card(\llbracket E\rrbracket ^\nu (\rho )) \, = \, 1$ we have $\{x'\} \, = \, \llbracket E\rrbracket ^\nu (\rho )$ and thus:

$\begin{array}{l} card(\llbracket E\rrbracket ^\nu (\rho ) \, = \, 1 \, ∧ \, \llbracket P\rrbracket ^\tau (\{\rho ⊕\lbb x\leadsto \llbracket E\rrbracket ^\nu \rbb\}) \, ≠ \, \{\} \\
 \,  \, ⇒ \\
\llbracket P\rrbracket ^\tau (\{\rho ⊕\lbb x\leadsto \llbracket E\rrbracket ^\nu \rbb\}) \, ≠ \, \{\} \\
\end{array}$

which is true by logic. 

\subsubsection{The soundness of the exists-elimination rule of FOPLN}

Our rule is:
 
 $\begin{array}{l} \displaystyle\frac{HYP, \, δ(α) \, ⊢ \, P[α/x] \, ⇒ \, Q, \,  \, HYP, \, δ(α) \, ⊢ \, ∃ \, x \, \bullet  \, P}{HYP \, ⊢ \, Q} \,  \, α \, \text{fresh,} \,  \, x∖Q \,  \,  \, \text{     ∃-elim} \\
\end{array}$

We will manipulate our rule into the best form for validation. First note that since we have $α$ fresh we know $α\setminus P$ which renders the $δ(α)$ in the second premise redundant.\footnote{Our motivation for including this redundant term is to provide symmetry when using the rule in inverted form for tree proofs.}   The $δ(α)$ in the first premise can be moved across the turnstile to appear on the LHS of an implication.  We can furthermore combine the two premises as a conjunction, and rewrite our inference rule as:

$\begin{array}{l} \displaystyle\frac{HYP \, ⊢ \, (δ(α) \, ⇒ \, (P[α/x]⇒Q)) \, ∧ \, (∃x\bullet P)}{HYP \, ⊢ \, Q} \,  \,  \,  \, α \, \text{fresh,} \,  \, x∖Q \,  \\
\end{array}$   

which by propositional logic results : $p⇒(q⇒r) \, ⇔ \, p∧q⇒r$ and $p⇒q \, ⇔ \, ¬p∨q$ can be written as:

 $\begin{array}{l} \displaystyle\frac{HYP \, ⊢ \, (¬(δ(α) \, ∧ \, (P[α/x]) \, ∨ \, Q)) \, ∧ \, (∃x\bullet P)}{HYP \, ⊢ \, Q} \,  \,  \,  \, α \, \text{fresh,} \,  \, x∖Q \,  \\
\end{array}$   

and to validate we show:

$\begin{array}{l} \llbracket  \, ¬(δ(α) \, ∧ \, P[α/x]) \, ∨ \, Q) \, ∧ \, (∃x\bullet P) \, \rrbracket  \\
 \,  \,  \, ⊆ \\
\llbracket  \, Q \, \rrbracket  \\
\end{array}$.

which, recalling the semantic rules

$\begin{array}{l} \llbracket ¬P\rrbracket ^\tau (\mathcal{E} ) \, = \, \mathcal{E}  \, \setminus \llbracket P\rrbracket ^(\mathcal{E} ) \\
\llbracket P∧Q\rrbracket ^(\mathcal{E} ) \, = \, \llbracket P\rrbracket (\mathcal{E} ) \, ∩ \, \llbracket Q\rrbracket ^\tau (\mathcal{E} ) \\
\llbracket P∨Q\rrbracket ^(\mathcal{E} ) \, = \, \llbracket P\rrbracket (\mathcal{E} ) \, ∪ \, \llbracket Q\rrbracket ^\tau (\mathcal{E} ) \,  \\
\end{array}$

we can express as:

$\begin{array}{l} ( \, ( \, \mathcal{E}  \, \setminus \llbracket (δ(α) \, ∧ \, P[α/x]\rrbracket  \, ) \, ∪ \, \llbracket Q\rrbracket ^\tau (\mathcal{E} ) \, ) \, ∩ \, \llbracket  \, (∃x\bullet P) \, \rrbracket  \\
 \,  \,  \, ⊆ \\
\llbracket  \, Q \, \rrbracket  \\
\end{array}$.

We first derive two lemmas.

Applying the semantic rule for $δ(E)$ to $δ(α)$ gives:

$\llbracket δ(α)\rrbracket ^\tau (\mathcal{E} ) \, = \, \{ \, \rho  \, | \, \rho  \, ∈ \, \mathcal{E}  \, ∧ \, card(\llbracket α\rrbracket )^\rho  \, = \, 1 \, \}$

Noting that from set theory $card(s) \, = \, 1 \, ⇔ \, (∃ \, x \, \bullet  \, s \, = \, \{x\})$ we can write this equiviently as:

$\llbracket δ(α)\rrbracket ^\tau (\mathcal{E} ) \, = \, \{ \, \rho  \, | \, \rho  \, ∈ \, \mathcal{E}  \, ∧ \, (∃ \, x' \, \bullet  \, \llbracket α\rrbracket ^(\rho ) \, = \, \{x'\}) \, \}$

But in our proof rule $α$ {\em is} an element, and we thus have:

 $\llbracket δ(α)\rrbracket ^\tau (\mathcal{E} ) \, = \, \mathcal{E} $

and therefore $\mathcal{E}  \, = \, \{ \, \rho  \, | \, \rho  \, ∈ \, \mathcal{E}  \, ∧ \, (∃ \, x' \, \bullet  \, \llbracket α\rrbracket ^(\rho ) \, = \, \{x'\}) \, \}$

from which we can conclude:

Lemma 1. $∃ \, x' \, \bullet  \, \llbracket α\rrbracket ^(\rho ) \, = \, \{x'\}$

Which we will use in the proof of:

Lemma 2. $\llbracket δ(α) \, ∧ \, P[α/x]\rrbracket ^(\mathcal{E} ) \, = \, \llbracket ∃x\bullet P\rrbracket ^\tau (\mathcal{E} )$

Proof of lemma 2.

We recall the following semantic rules:

$\begin{array}{l} \llbracket P[E/x]\rrbracket ^\tau (\mathcal{E} ) \, = \, \{\rho  \, | \, \rho  \, ∈ \, \mathcal{E}  \, ∧ \, \llbracket P\rrbracket ^(\{\rho ⊕\lbb x\leadsto \llbracket E\rrbracket \rbb\})\} \\
\llbracket ∃x\bullet P\rrbracket ^\tau (\mathcal{E} ) \, = \, \{ \, \rho  \, | \, \rho  \, ∈ \, \mathcal{E}  \, ∧ \, (∃ \, x' \, \bullet  \, \llbracket P\rrbracket ^\tau (\{\rho ⊕\lbb x\leadsto \{x'\}\rbb\}) \, ≠ \, \{\}) \, \} \\
\end{array}$

Thus, along with the derived semantics for $δ(α)$ given above, we have:

$\begin{array}{l} \llbracket δ(α) \, ∧ \, P[α/x]\rrbracket ^(\mathcal{E} ) \, = \,  \\
 \,  \,  \,  \,  \,  \, \{ \, \rho  \, | \, \rho  \, ∈ \, \mathcal{E}  \, ∧ \, (∃x' \, \bullet  \, \llbracket α\rrbracket ^(\rho ) \, = \, \{x'\}) \, ∧ \, \llbracket P\rrbracket ^\tau (\{\rho ⊕\lbb x\leadsto \llbracket α\rrbracket ^\nu (\rho )\rbb\}) \, ≠ \, \{\} \, \} \\
= \, \text{“extending the scope of }x'\text{ to the right and writing }\{x'\}\text{ for }\llbracket α\rrbracket ^\nu (\rho )\text{”} \\
\{ \, \rho  \, | \, \rho  \, ∈ \, \mathcal{E}  \, ∧ \, (∃x' \, \bullet  \, \llbracket α\rrbracket ^(\rho ) \, = \, \{x'\} \, ∧ \, \llbracket P\rrbracket ^\tau (\{\rho ⊕\lbb x\leadsto \{x'\}\}) \, ≠ \, \{\}) \, \} \\
= \, \text{by lemma 1} \\
\{ \, \rho  \, | \, \rho  \, ∈ \, \mathcal{E}  \, ∧ \, (∃x' \, \bullet  \, \llbracket P\rrbracket ^\tau (\{\rho ⊕\lbb x\leadsto \{x'\}\}) \, ≠ \, \{\}) \, \} \\
= \, \text{“see denotational rule for }∃x\bullet P\text{ quoted above”} \\
\llbracket ∃x\bullet P\rrbracket ^\tau (\mathcal{E} ) \\
\end{array}$

Which completes the proof of lemma 2.

Returning to the validation of the ∃-elim rule, we recall that we have to show:

$((\mathcal{E} ∖\llbracket δ(α)∧P[α/x]\rrbracket ^(\mathcal{E} )) \, ∪ \, \llbracket Q\rrbracket ^\tau (\mathcal{E} )) \, ∩ \, \llbracket ∃x\bullet P\rrbracket ^\tau (\mathcal{E} ) \,  \, ⊆ \,  \, \llbracket Q\rrbracket ^\tau (\mathcal{E} )$

we now abstract with $\mathcal{E} $ as $A$, $\llbracket δ(α)∧P[α/x]\rrbracket ^(\mathcal{E} )$ as $B$,$\llbracket ∃x\bullet P\rrbracket ^\tau (\mathcal{E} )$ as $B'$ and $\llbracket Q\rrbracket ^\tau (\mathcal{E} )$ as $C$.

We must show $( \, (A \, \setminus B) \, ∪ \, C \, ) \, ∩ \, B' \,  \, ⊆ \,  \, C$

i.e. by distribution of set intersection over set union:

$( \, (A∖B) \, ∩ \, B' \, ) \,  \, ∪ \,  \, (C∩B') \,  \, ⊆ \, C$ 

But from lemma 2 we have $B=B'$ and from set theory we have:

$B=B' \, ⇒ \, (A∖B') \, ∩ \, B \, = \, \{\}$

Thus our task becomes to show

$\{\} \, ∪ \, (C∩B') \, ⊆ \, C$

Which follows directly.

\section{Conservative extensions to bunch notation}

So far the only notation that distinguishes our axiomatic bunch theory from the axiomatic set theory on which it is based, is the unpacking operator $\unpack $.

However, we can now extend our notations through a series of definitions. We assume standard definitions from set theory, e.g. for set union, intersection, and difference. We will also use set comprehension expressions of the form \(\{ \, x \, | \, P \, \bullet  \, E \, \}\), representing the set of values expression $E$ can take as $x$ ranges over the values allowed by predicate $P$.

We define the $null$ bunch, bunch union and intersection, and bunch inclusion:

\( \, null \,  \, \defs  \,  \, \unpack \{ \, \} \, \)

\( \, E \, , \, F \, \defs  \, \unpack (\{E\} \, ∪ \, \{F\}) \, \)

\( \, E \, ‘ \, F \, \defs  \, \unpack (\{E\} \, ∩ \, \{F\}) \, \)

\( \, E \, : \, F \, \defs  \, \{E\}⊆\{F\} \, \)

Now conditional expressions:

\( \, \IF \,  \, g \,  \, \THEN \,  \, E \,  \, \ELSE \,  \, F \,  \, \END \,  \, \defs  \,  \,  \, g\bguard  \, E \, , \, ¬g\bguard  \, F \, \)

And a form of bunch comprehension:

\( \, ∮ \, x \, | \, x∈S \, ∧ \, P \,  \, \bullet  \, E \,  \, \defs  \,  \,  \, \unpack \{ \, x \, | \, x∈S \, ∧ \, P \, \bullet  \, E\} \, \)

Since $y \, ∈ \, \{ \, x \, | \, x∈S \, ∧ \, P \,  \, \bullet  \, E \, \} \,  \, ⇔ \,  \,  \, ∃ \, x \, \bullet  \, x∈S∧P \, ∧ \, y:E$ we can
write $\{ \, x \, | \, x∈S \, ∧ \, P \,  \, \bullet  \, E \, \}$ as $\{ \, x \, | \, true \, \bullet  \, x∈S \, ∧ \, P \, \bguard  \,  \, E \, \}$ which we abbreviate to $\{ \, x \, \bullet  \, x∈S \, ∧ \, P \, \bguard  \,  \, E \, \}$. Also, since $ \, x∈S \, ∧ \, P \, \bguard  \,  \, E$ is an expression, we can take $\{ \, x \, \bullet  \, E \, \}$ as a general form for set comprehension, with the proviso that $E$ must imply the type of $x$. Corresponding to this we can use $∮x \, \bullet  \, E$ as a general form for bunch comprehension.

We now come to some redefinitions, required to lift existing definitions, which apply only to elements, to apply element-wise to bunches. 

We incur a proof obligation to show the new definitions will be equivalent to the old when applied to arguments for which the old definitions were defined, i.e. to elements. To avoid need for renaming this obligation can be discharged prior to the redefinition. Thus, noting that when $E$ and $S$ are elements:

\( \, \{E\} \, ⊆ \, ⋂ \, \{ \, S \, \} \,  \, ⇔ \,  \, E \, ∈ \, S \, \)

we are now free to introduce the following redefinition of $E \, ∈ \, S$, in which $E$ and $S$ are not restricted to being elements, because we know it is compatible with the existing definition in the case where $E$ and $S$ are elements.

\( \, E \, ∈ \, S \, \defs  \,  \, \{E\} \, ⊆ \, ⋂ \, \{ \, S \, \} \, \)

The effect of this definition is that for $E \, ∈ \, S$ to hold every $e:E$ must be a member of every $s:S$. This is in line with our use of bunches to represent non-deterministic terms, and guarantees that membership will hold after the non-determinism is resolved.

Here is a redefinition which lifts unpacking operators. The new definition of $\unpack $ is expressed in terms of the original $\unpack $, which appears in the body of the original definition. 

\( \, \unpack E \, \defs  \, \unpack (∪\{E\}) \, \).

A more general way of lifting an operation, and one where the accompanying proof obligation can be safely assumed to be discharged, is to represent it in terms of a bunch comprehension.  Here is the definition of the lifted power set operator:

\( \, \pow (S) \,  \, \defs  \,  \, ∮s \, \bullet  \, s:S \, \bguard  \, \pow (s) \, \)

\section{Some results in bunch theory}

\subsection{Bound variables in bunch theory range over elements} 

This result follows directly from FOPLN, the underlying logic for our bunch  theory \cite{arXivLogic}.

\subsection{Properties of bunch union and bunch intersection}

The following are immediately provable by appealing to  the definitions of $null$, bunch intersection, and bunch union. 

$\begin{array}{lll} E \, ‘ \, F \, = \, F \, ‘ \, E \,  \,  \,  \, & \,  \,  \, E \, , \, F \, = \, F \, , \, E \,  \,  \,  \,  \, & \,  \,  \,  \,  \, \text{commutativity} \\
(E \, ‘ \, F) \, ‘ \, G \, = \, E \, ‘ \, (F \, ‘ \, G) \,  \,  \,  \, & \,  \,  \, E \, , \, (F \, , \, G) \, = \, (E \, , \, F) \, , \, G \,  \, & \, \text{associativity} \\
E \, ‘ \, (F \, , \, G) \, = \, (E \, ‘ \, F) \, , \, (E \, ‘ \, G) \,  \,  \, & \,  \, E \, , \, (F \, ‘ \, G) \, = \,  \, ( \, E, \, F) \, ‘ \, (E \, , \, G) \,  \, & \, \text{distributivity} \\
E \, , \, (F \, , \, G) \, = \, (E \, , \, F) \, , \, (E \, , \, G) \,  \,  \, & \,  \, E \, ‘ \, (F \, ‘ \, G) \, = \, (E \, ‘ \, F) \, ‘ \, (E \, ‘ \, G) \,  \,  \, & \, \text{distributivity} \,  \\
E \, : \, E \, , \, F \, & \,  \, E \, ‘ \, F \, : \, E \,  \, & \,  \, \text{inclusion} \\
E \, , \, null \, = \, E \,  \, & \,  \, E \, ‘ \, null \, = \, null \,  \, & \,  \, \text{zero and unit} \\
E \, , \, E \, = \, E \,  \, & \,  \, E \, ‘ \, E \, = \, E \, & \, \text{idempotence} \\
\end{array}$

\subsection{Equality properties}

The set equality axiom $∀ \, x \, \bullet  \, x∈A⇔x∈B \, ⇒ \, A=B$ now requires the proviso that $A$ and $B$ are elements, but we have the following laws for equality of general bunch expressions $E$ and $F$ 

$\begin{array}{l} ∀ \, x \, \bullet  \, x \, : \, E \, ⇔ \, x \, : \, F \,  \, ⇒ \,  \, E=F \\
\{E\} \, = \, \{F\} \,  \, ⇔ \,  \, E=F \\
\end{array}$

\subsection{Properties of guarded bunches}

The guarded bunch has distributive properties which can be obviously proved by case analysis:

$\begin{array}{l} g\bguard  \, E \, , \, F \,  \,  \, = \,  \,  \, (g\bguard  \, E) \, , \, (g\bguard  \, F) \\
g\bguard  \, E \, ‘ \, F \,  \,  \,  \, = \,  \,  \, (g\bguard  \, E) \, ‘ \, (g\bguard  \, F) \\
\end{array}$

\subsection{Generalisation of certain laws}

We have the following “part of” and “member of” laws: 

$\begin{array}{l} x \, : \, E,F \,  \,  \, ⇔ \,  \,  \, x:E \, ∨ \, x:F \,  \,  \, \text{($x$ an element)} \\
x \, ∈ \, E \, F \,  \,  \, ⇔ \,  \,  \, x∈E \, ∧ \,  \, x∈F \\
\end{array}$

Which generalise to the following bunch comprehension laws:

$\begin{array}{l} x \, : \, ∮ \, y \, \bullet  \, y \, : \, Y \, \bguard  \, E \,  \,  \, ⇔ \,  \, ∃ \, y \, \bullet  \, y:Y \, ∧ \, x:E \,  \,  \,  \,  \, (x \,  \, \text{an element}) \\
x \, ∈ \, ∮ \, y \, \bullet  \, y \, : \, Y \, \bguard  \, E \,  \,  \, ⇔ \,  \, ∀ \, y \, \bullet  \, y:Y \, ⇒ \, x∈E \,  \,  \,  \,  \,  \\
\end{array}$

Some of the set theory axioms can be generalised. The ordered pair axiom  $E\mapsto F \, ∈ \, (s \, × \, t) \, ⇔ \, E \, ∈ \, s \, ∧ \, F \, ∈ \, t$ holds for non-elemental values, and this has already been validated in our model. 

The power set axiom $s \, ∈ \, \pow (t) \,  \, ⇔ \,  \, (∀ \, x \, \bullet  \, x \, ∈ \, s \, ⇒ \, x \, ∈ \, t)$ where $x$ is not free in $s$ or $t$, still holds when $t$ is a non-elemental value.

\section{Axioms and semantics of “Improper bunch theory”}

For modelling program semantics within a total correctness framework we associate with each type $T$ an `improper' bunch $⊥ \, _T$. We consider it to be of type $T$. We require the following additional axioms.

$\begin{array}{l}  \\
E \, : \, ⊥ \,  \,  \, \text{“maximality”} \end{array}$

$\begin{array}{l}
⊥ \, : \, A,B \,  \, ⇒ \,  \, A=⊥ \,  \, ∨ \,  \, B=⊥ \,  \,  \, \text{“atomicity”} \end{array}$

$\begin{array}{l}
\{⊥ \, _T\} \,  \, = \,  \, ⊥ \, _{\pow (T)} \,  \,  \, \text{“improper packaging”} \end{array}$

$\begin{array}{l}
\unpack ⊥ \, _{\pow (T)} \,  \, = \, ⊥ \, _T \,  \,  \,  \,  \, \text{“improper unpackaging”} \end{array}$

$\begin{array}{l}
\end{array}$ 

And we need to modify the axiom of proper bunch theory that says $\{E\}$ is an element, which we can now assert only when $E \, ≠ \, ⊥ \, $

$\begin{array}{l} E \, ≠ \, ⊥ \,  \, ⇒ \,  \, δ( \, \{E\} \, ) \,  \, \text{“guarded element”} \\
\end{array}$

The maximality axiom tells us that for any type, any bunch other than the improper bunch is a proper sub-bunch of the improper bunch. 

The atomicity axiom tells us that $⊥$ cannot be constructed from the union of two bunches both different from itself.

The packaging axioms tell us that impropriety cannot be removed by packaging or unpackaging, although the results of packing and unpacking an improper bunch will be of the expected type.

Note that we no longer have an unguarded axiom to say $\{E\}$ is an element,  but the {\em atomicity} of $\{E\}$ follows from the atomicity axiom together with the guarded element axiom.
 
\subsection{Extending our model to include the improper bunch}

Within our semantic universe we distinguish sets constructed directly from $BIG$, which we term “basic” or “given” sets, from those defined by using the power-set and set product constructors.

We recall once more that any such given set is maximal and corresponds to a type, as do all sets constructed from maximal sets.

The semantic support for a basic set, $T$, say, in our language of improper bunch theory, is given by a set $T'$ in the meta-language of our semantic universe which contains the elements of the set $T$, along with an additional distinguished element $κ_T$.

We define the semantic representation of $⊥_T$ by $\llbracket ⊥_T\rrbracket  \, = \, T'$

Thus if the given set $T$ is defined so that $T=\{a,b\}$ then to support this within our semantic universe we have a set $T'=\{a,b,κ_T\}$, and we have $\llbracket ⊥_T\rrbracket  \, = \, \{a,b,κ_T\}$.

The denotations of all possible bunches of type $T$ are as follows: 

\[⟦null_T\rrbracket =\{\}, \, \llbracket a\rrbracket =\{a\}, \, \llbracket b\rrbracket =\{b\}, \, \llbracket a,b\rrbracket =\{a,b\}, \, \llbracket ⊥_T\rrbracket =\{a,b,κ_T\} \, \]

We note that $⊥_T$ is the only term of type $T$ whose denotation
contains the distinguished element $κ_T$. This allows us to model the atomicity of the improper bunch. We also note that the denotation of the improper bunch for a given type is a maximal set. This allows us to model the maximality of the improper bunch.

We use similar techniques for any constructed type. 

Within each maximal constructed set of the semantic universe, we have a $κ$ element defined structurally as: $κ_{T×U} \, = \, κ_T \, \mapsto  \, κ_U$ and $κ_{\pow (T)} \, = \, \llbracket ⊥_T\rrbracket $. 

Denotations for the improper bunch of each type are:

$\begin{array}{l} \llbracket ⊥_T\rrbracket  \, = \, T' \, \text{($T$ a given set)} \\
\llbracket ⊥_{T×U}\rrbracket  \, = \, \llbracket ⊥_T\rrbracket  \, × \, \llbracket ⊥_U\rrbracket  \\
\llbracket ⊥_{\pow (T)}\rrbracket  \, = \, \pow (\llbracket ⊥_T\rrbracket ) \\
\end{array}$

Thus, the denotation for the improper bunch of any type will be a maximal set.

We recall that the purpose of the improper bunch is to represent the effect of running a program outside its pre-condition. In line with the philosophy of total correctness, we take a pessimistic view of such effects, which may go well beyond non-termination, and include, for example, stack and array overflows which corrupt pointers or code, leading to totally unpredictable results - the car may crash! To reflect this we want the improper bunch to be absorptive with respect to all expression connectives. We extend a number of definitions to take account of special cases arising from this property. We can do this using bunch guard. In each of the following definitions, the definition body is a bunch of terms, the first of which is a guarded bunch. If the bunch guard is true, that term, and hence the whole body of the definition,  represents an improper bunch. Otherwise the guarded term represents $null$ and disappears.\footnote{It is possible to write denotations for these forms directly, rather than using redefinition. For example $\llbracket \{E\}\rrbracket ^\nu (\rho ) \, \defs  \, \IF \, \llbracket E\rrbracket ^\nu (\rho ) \, = \, \llbracket ⊥_T\rrbracket  \,  \, \THEN \, \llbracket ⊥ \, _{\pow (T)}\rrbracket  \, \ELSE \, \{ \, \llbracket E\rrbracket ^\nu (\rho ) \, \} \, \END$.}

$\begin{array}{l} \{E\} \,  \,  \, \defs  \,  \,  \,  \, E=⊥ \, _T \, \bguard  \, ⊥ \, _{\pow (T)} \,  \,  \, , \,  \,  \, \{E\} \end{array}$

$\begin{array}{l}
\unpack E \,  \,  \, \defs  \,  \,  \, E=⊥ \, _{\pow (T)} \, \bguard  \, ⊥ \, _T \,  \,  \,  \, , \,  \,  \, \unpack E \end{array}$

$\begin{array}{l}
A∪B \,  \,  \, \defs  \,  \,  \, A=⊥ \, ∨ \, B=⊥ \,  \, \bguard  \,  \, ⊥ \,  \,  \, , \,  \,  \, A \, ∪ \, B \end{array}$

$\begin{array}{l}
A∩B \,  \,  \, \defs  \,  \,  \,  \, A=⊥ \, ∨ \, B=⊥ \,  \, \bguard  \,  \, ⊥ \,  \,  \, , \,  \,  \, A \, ∩ \, B \end{array}$

$\begin{array}{l}
D \, \mapsto  \, E \,  \, \defs  \,  \, D=⊥ \, ∨ \, E=⊥ \,  \, \bguard  \,  \, ⊥ \,  \,  \, , \,  \, D \, \mapsto  \, E \end{array}$

$\begin{array}{l}
S \, × \, T \,  \, \defs  \,  \,  \, S=⊥ \, ∨ \, T=⊥ \,  \, \bguard  \,  \, ⊥ \,  \, , \,  \, S \, × \, T \\
\end{array}$

For bunch union and bunch intersection we can deduce absorptive properties from the above, e.g. 

$\begin{array}{l} A \, , \, ⊥_T \,  \, \eq \,  \, \text{“by definition of bunch union”} \\
\unpack (\{A\} \, ∪ \, \{⊥_T\}) \,  \, \eq \,  \, \text{“by improper packaging axiom”} \\
\unpack (\{A\} \, ∪ \, ⊥_{\pow (T)} \, ) \,  \, \eq \,  \, \text{“by guarded redefinition of set union”} \\
\unpack ⊥_{\pow (T)} \,  \, \eq \, \text{“by improper unpackaging axiom”} \\
⊥_T \,  \\
\end{array}$ 

One construct that is not subject to these absorptive properties is the bunch guard, since the second guarded bunch axiom tells us that $false \, \bguard  \, E \, = \, null$. This is to capture the effect, in the world of values, of code that cannot run, e.g. the contribution from the branch of a conditional statement that is not executed.

\subsection{Validation of the axioms of improper bunch theory}

Let ${\cal \, D}(T)$ be the set of denotations for bunches of type $T$. We formalise the properties, already discussed above, that support maximality and atomicity of the improper bunch. 

Property. $x \, ∈ \, {\cal \, D}(T) \, ⇒ \, x \, ⊆ \, \llbracket ⊥_T\rrbracket $

To validate of the maximality of the improper bunch, $E \, : \, ⊥$, we must show $\llbracket E \, : \, ⊥\rrbracket ^\tau (\mathcal{E} ) \, = \, \mathcal{E} $. We need the property, already discussed above $x \, ∈ \, {\cal \, D}(T) \, ⇒ \, x \, ⊆ \, \llbracket ⊥_T\rrbracket $

$\begin{array}{l} \llbracket E \, : \, ⊥\rrbracket ^\tau (\mathcal{E} ) \, = \, \mathcal{E}  \,  \,  \,  \,  \, \eq \,  \, \text{“by semantics of bunch inclusion”} \\
\{ \, \rho  \, | \, \rho  \, ∈ \, \mathcal{E}  \,  \, ∧ \,  \, \llbracket E\rrbracket ^\nu (\rho ) \, ⊆ \llbracket ⊥\rrbracket  \, \} \,  \, = \, \mathcal{E}  \,  \, \eq \,  \, \text{“for arbitrary } \, \rho  \, \text{”} \\
\llbracket E\rrbracket ^\nu (\rho ) \, ⊆ \llbracket ⊥\rrbracket  \,  \, \eq \,  \, \text{“by property given above”} \\
true \, \square  \,  \\
\end{array}$

Validation of the axioms for improper packaging, improper unpacking, and guarded element follow immediately from the revised semantics of packaging.

For a fuller treatment which includes validation of the atomicity of the improper bunch see \cite{arXivBunches_v2}

\section{Related work}

The main material on bunch theory and its applications is found in Hehner's book “A practical theory of programming” \cite{ECRH93}, which uses a predicative style for program descriptions. Bunches are used to incorporate non-determinism in functions by Morris and Bunkenburg \cite{MB01}. They propose a model that axiomatises boolean and function types in the presence of non-determinism and considers the adaptation of logic to bunches of truth values. This contrasts with our approach which provides a full integration with set theory and isolates the use of bunches  from logic by distinguishing truth semantics from value semantics, so we can continue to use classical predicate logic along with bunches.

An early application of bunches was to formal grammars, with a language being a bunch. In “The functional treatment of parsing” by René Leermakers \cite{RL93} (1993, second edition 2012) non-deterministic functional parsing algorithms are used with the aim of bridging the gap between the parsing techniques of computer science and of natural language analysis. The author notes that the use of bunches smooths the transition from non-deterministic to deterministic algorithms, and that the use of bunches allows the simple expression of formulae which “blow up” if expressed in set theory. We illustrated the same point in section \ref{simplicity}

\section{Conclusions}

Set Theory is foundational in mathematics. A set is a packaged collection, and when packaging and collecting are separated, we obtain Bunch Theory. This is a radically different theory with significant advantages.

In the area of “descriptions”, traditional approaches give rise to the possibility of undefined or non-denoting terms, e.g. when a partial function is applied outside its domain. This is a problem that has generated an enormous amount of discussion, from the early approaches of Russell and Frege, to the treatment of undefined terms in specification languages. The essential problem is that no approach (apart from Bunch Theory) is able to include “nothing” as a conceptual object.  In our theory, undefined terms still denote, and are handled comfortably.

Bunches, along with the convention of “lifting” most operations and function application, naturally represent non-deterministic terms. We illustrate this by describing a semantics of “prospective values”, in which $S\diamond E$ represents the value or values expression $E$ will take after program $S$ has executed. The simplicity of the semantics we derive above depends on the {\em preservation of homogeneity} when non-determinism is introduced: the type of $S \, \diamond  \, E$ is the same as the type of $E$. This is essential to obtain the simplicity of the rule $S\asemi T \, \diamond  \, E \,  \, \eq \,  \, S \, \diamond  \, (T \, \diamond  \, E)$. \footnote{If we used sets instead of bunches to record the possible values of $E$ after $S$. Then $S \, \diamond  \, E$ would be a set, and $S \, \diamond  \, T \, \diamond  \, E$ would be a set of sets, and our rule would need to incorporate generalised set union to remove the build up of unwanted structure.}  

Our other illustrations show how bunch theory can provide continuations,  preferential choice that takes account of the feasibility of continuations in a backtracking context, and the integration of probabilistic choice. We introduce an “improper bunch”, which is maximally non-deterministic, to represent failed computations. Later in the paper, after “wholistic” function application is introduced, we use bunch theory to  represent, in a notationally pleasant way, a formal grammar, composed, mathematically, of simultaneous fixed point equations which are shown to have solutions. 

Our formal description of bunch theory is constructed by adding packing and unpacking operations to a typed set theory. We use a theory, taken from the B method, which distinguishes predicates, which are subject to proof, from expressions, which are subject to evaluated. This separation means that adding bunches to our theory leaves the underlying classical logic unchanged. This contrasts with the approaches of Hehner and Morris, who both have to contend with additional truth values in their logics. 

We provide and validate a model for our theory. This allows us to see that the axiomatisation of bunches is orthogonal to the existing axiomatisation of set theory. Our method could be used to add bunches to other axiomatisations of set theory.

The null bunch, which conceptualises nothing, is used again and again in our formulations: for results from computations which do not occur (e.g. the $\ELSE$ branch of a conditional statement when the $\IF$ branch is selected), for infeasible continuations during backtracking search, for the parts of conditional expressions which are eliminated when the expression guard is evaluated, and for nonsensical descriptions (the current King of France) arising when partial functions are applied outside their domains. Our experience working with bunches makes a theory that is unable to conceptualise “nothing” seem as defective as a theory of arithmetic that omits zero.

\bibliography{stoddart,fa,rev}

\bibliographystyle{alpha}

\vspace{2 em}

\appendix

{\bf \huge Appendices}

\section{Prospective values in a fixed point treatment of loops}

\label{Fixpoint}

We have an iteration construct

$\WHILE \, g \, \DO \,  \, S \, \END$, which  we will refer to  as $W$.

We define $W$ recursively by unwrapping the first iteration of  the  loop:

$W \, \eq \, \IF \, g \, \THEN \, S \, \asemi  \, W \, \ELSE \, skip \, \END$

and for our prospective value approach we  are  interested in obtaining a recursive equation which we can eventually  use to establish the existence of $W\diamond E$, the bunch of prospective values for expression $E$ following $W$.

We first establish the monotonicity property, namely $E \, : \, F \, ⇒ \, (S\diamond  \, E) \, : \, (S\diamond F)$, which will hold for any program $S$ defined using the semantic primitives of section \ref{primitives}. We start from the monotonicity property of WP semantics that for any program $S$ and predicates $P,Q$, that $(P⇒Q) \, ⇒ \, ([S]P \, ⇒ \, [S]Q$. This can be re-expressed in CWP terms as  $(P⇒Q) \, ⇒ \, (⟨S⟩P \, ⇒ \, ⟨S⟩Q)$. I.e. if $S$ might establish $P$ and $P⇒Q$, then $S$ might establish $Q$. Now let $E$ and $F$ be bunch expressions such that $E \, : \, F$. Let $z$ be an arbitrary element of  the correct  type, then $z \, : \, E \, ⇒ \, z \, : \, F$ and thus using the monotonicity property for CWP we have  $(z \, : \, E \, ⇒ \, z \, : \, F) \, ⇒ \, (⟨S⟩z \, : \, E \, ⇒ \, ⟨S⟩z \, : \, F)$. We can now argue as follows:

$\begin{array}{l} (z \, : \, E \, ⇒ \, z \, : \, F) \, ⇒ \, (⟨S⟩z \, : \, E \, ⇒ \, ⟨S⟩z \, : \, F) \, \eq \, \text{“basic law from section \ref{basiclaw}”} \\
(z \, : \, E \, ⇒ \, z \, : \, F) \, ⇒ \, (z \, : \, (S\diamond E) \, ⇒ \, ( \, z \, : \, (S\diamond E) \, ⇒ \, z \, : \, (S\diamond F) \,  \\
\eq \, `` property \, of \, bunch \, inclusion\rightq  \\
E \, : \, F \, ⇒ \, (S\diamond E) \, : \, (S\diamond F) \\
\end{array}$

We will now be able to derive a fixed point equation in expression which are monotonic wrt bunch inclusion. The ordering properties of bunch inclusion are similar to those of set inclusion, but we cannot think of bunches as forming a partial order, and in particular a lattice, because we cannot package them in a set without  them losing their identity.

Let us first establish, therefore, that a recursive equation in bunch expressions will nevertheless give us  the basis to call on Tarski's fixed point theorem for complete lattices, namely:
 
{\bf Theorem} If the elements of a set $P$ form a complete lattice under some partial order and $F \, ∈ \, P \, → \, P$ is monotonic wrt to that  order, then the equation $X \, = \, F(X)$ has a minimum solution (a minimum fixed point).

We must show we can call on this theorem when given a recursive equation in bunch {\em expressions}, $U \, = \, E$, where $U$ occurs as a sub-expression in $E$, without variable capture, and $E$ is monotonic increasing wrt $U$ and bunch inclusion ordering. Under the assumption $\unpack \{U\} \, = \, U$ We can rewrite our equation as a fixed point equation in set expressions  

$\{U\} \, = \, \{E[ \, \unpack \{U\} \, / \, U]\}$ 

or writing  $X$ for $\{U\}$  and $D$ for $\{E\}$ we obtain the following equation in which $D$ will be monotonic increasing wrt $X$ and inverse set inclusion ordering:

 $X \, = \, D[ \, \unpack X \, / \, U \, ]$ 

Which using lambda abstraction we can write as: 

$X \, = \, (λ \, Y \, \bullet  \, D \, [ \, \unpack Y \, / \, U \, ])X$

Giving us a classical fixed point equation with a set inclusion ordering that forms a complete lattice, having the form $X=F(X)$ where $F \, ∈ \, \pow (T) \, → \, \pow (T)$ for some $T$ with the associated partial order $(\pow (T),⊇)$ which is a complete lattice. 

The above conversion fails, however, when the assumption  $\unpack \{U\} \, = \, U$ does not hold, and this will occur when $U=⊥_T$, since $\{⊥_T\} \, = \, ⊥_{\pow (T)}$. We will still be able to provide a full analysis of $W\diamond E$, but we  will need to first introduce a weaker prospective value semantics in which the improper bunch does not occur, because the effect of a non-terminating computation is to yield the prospective value $All$, i.e. the most non-deterministic proper bunch, rather than $⊥$ . We will call this "liberal prospective value" or LPV,  semantics and it will have the same discriminating power as weakest liberal pre-condition (WLP) semantics: neither is able to discriminate between a non-terminating computation and a maximally non-deterministic computation that does terminate.

Let us write the LPV value of expression $E$ after computation $S$ as $S \, \diamond _0{} \, E$.  The rules for $S \, \diamond _0{} \, E$ over our program connectives are identical to those for $S \, \diamond  \, E$ with the single exception of the pre-conditioned computation $P \, | \, S$, where we  have

$P|S \, \diamond _0{} \, E \,  \, \eq \,  \,  \, ¬P \, \bguard  \, All \, , \,  \, S \, \diamond _0{} \, E$

i.e. a computation carried out outside its precondition has the prospective value $All$.

We now return to our analysis of the loop $W$, which we defined as

$W \, \eq \, \IF \, g \, \THEN \, S \, \asemi  \, W \, \ELSE \, skip \, \END$

so in terms of LPV semantics

$W \, \diamond _0{} \, E \,  \, \eq \,  \, \IF \, g \, \THEN \, S \, \asemi  \, W \, \ELSE \, skip \, \END \,  \, \diamond _0{} \,  \, E$

Applying our definition for the $\IF$ construct this becomes:

$\begin{array}{l} W \, \diamond _0{} \, E \,  \, \eq \,  \, g \, \Guard S \, \asemi  \, W \,  \, ⊓ \,  \, ¬g \, \Guard skip \,  \, \diamond _0{} \,  \, E \\
\eq \, `` \text{applying rule for choice}\rightq  \\
( \, g \, \Guard S \, \asemi  \, W \, \diamond _0{} \, E \, ) \,  \,  \, , \,  \, ( \, ¬g \, \Guard skip \,  \, \diamond _0{} \,  \, E \, ) \\
\eq \, `` \text{applying rule for sequential composition}\rightq  \,  \\
( \, g \, \Guard S \, \diamond _0{} \, W \, \diamond _0{} \, E \, ) \,  \,  \, , \,  \, ( \, ¬g \, \Guard skip \,  \, \diamond _0{} \,  \, E \, ) \\
\eq \, `` \text{applying rules for guard and skip}``  \\
( \, g \, \bguard  \, S \, \diamond _0{} \, W \, \diamond _0{} \, E \, ) \,  \,  \, , \,  \, ( \, ¬g \, \bguard  \, E \, ) \\
\end{array}$

Which, noting that $\diamond _0{}$ is right associative,  gives us a recursive equation in the bunch expression $W \, \diamond _0{} \, E$ :

$W \, \diamond _0{} \, E \,  \, \eq \, ( \, g \, \bguard  \, S \, \diamond _0{} \, W \, \diamond _0{} \, E \, ) \,  \,  \, , \,  \, ( \, ¬g \, \bguard  \, E \, )$

and we can conclude that this equation has a minimum solution for $W \, \diamond _0{} \, E$

Aside: For the non-terminating loop $\WHILE \, true \, \DO \, skip \, \END$ our equation for $W \, \diamond _0{} \, E$ is:

$\begin{array}{l} W \, \diamond _0{} \, E \, \eq \, (true \, \bguard skip \, \diamond _0{} \, W \, \diamond _0{} \, E) \, , \, false \, \bguard  \, E \\
`` \text{which reduces to}\rightq  \\
W \, \diamond _0{} \, E \, \eq \, W \, \diamond _0{} \, E \,  \\
\end{array}$

an equation which has every value as a fixed point, and in particular has the bunch $All$ as its least fixed point. End Aside

It remains to obtain the value of $W \, \diamond  \, E$, i.e. the prospective value for $E$ after $W$ rather than its liberal prospective value.  For this we note that although LPV cannot always distinguish between non-terminating computations and terminating computations that are maximally non-deterministic, it can do so for feasible loops, since a terminating loop will leave the loop guard false, and thus cannot be maximally non-deterministic. This enables us to express $W \, \diamond  \, E$ in terms of $W \, \diamond _0{} \, E$ as:

$W \, \diamond  \, E \,  \, \eq \,  \,  \, (W\diamond _0{}E) \, = \, All \,  \, \bguard  \, ⊥ \,  \,  \, , \,  \, (W \, \diamond _0{} \, E)$

\section{Precedence and associativity}

\label{precedence}

The following symbols are right associative: $\bullet  \, ⇒ \,  \, \diamond  \,  \, ∇ \,  \, ≡>$. 
All other binary connectives are left associative.

In descending order, with connectives of equal precedence included within brackets, e.g. $(* \, / \, mod)$, the order of precedence is:

$\begin{array}{l} \text{function application (symbol elided)} \,  \,  \, . \,  \\
(* \, / \, mod) \, (+ \, -) \\
∩ \, ∪ \, \bigoplus  \, → \, (, \, ') \,  \\
(> \, < \, ≤ \, ≥) \, (= \, ≠) \, ( \, : \,  \, ⊆ \, ⊂ \,  \, ∈ \, ∉ \, ) \, ∧ \, ∨ \, ⇒ \,  \, ⇔ \,  \\
\bullet  \, \text{as used in quantifications} \\
\bguard  \, \bpre  \\
:= \, \Guard ⊓ \, ⊔ \, ⟩⟩ \, ⊞ \, \asemi  \, | \, ( \, \diamond  \, ∇ \, ) \, ⊑ \\
( \, \defs  \, \eq \, ≡ \,  \,  \, ≡> \, ) \,  \\
\end{array}$

We have the following unary prefix symbols, whose precedence is above that of binary connectives. 

\( \, - \, \unpack  \, \pow  \, ∮ \, ¬ \, ∀ \, ∃ \, \)

The precedence level of $\bullet $, together with the definition of $∀$ and $∃$ as unary prefixes, defines the scope of quantified variables in predicates.

\end{document}